\documentclass{aa}
\usepackage{graphicx}
\usepackage{txfonts}

\begin{document}

\title{Effect of asymmetry of the radio source distribution on the
apparent proper motion kinematic analysis}
\author{Oleg Titov \inst{1} \and Zinovy Malkin \inst{2}}
\offprints{Oleg Titov, \email{Oleg.Titov@ga.gov.au}}
\institute{Geoscience Australia, GPO Box 378, Canberra, ACT 2601, Australia  \and
Central Astronomical Observatory at Pulkovo of RAS, Pulkovskoe~Ch. 65,
St.~Petersburg 196140, Russia}
\date{Received / Accepted}
\titlerunning{Effect of asymmetry of the radio source distribution}
\authorrunning{O. Titov \and Z. Malkin}

\abstract
{Information on physical characteristics of astrometric radio sources,
magnitude and redshift in the first place, is of great importance for
many astronomical studies.  However, data usually used in radio
astrometry is incomplete and often outdated.}
{Our purpose is to study the optical characteristics of more than 4000
radio sources observed by the astrometric VLBI technique since 1979.
Also we studied an effect of the asymmetry in the distribution of the
reference radio sources on the correlation matrices between vector
spherical harmonics of the first and second degrees.}
{The radio source characteristics were mainly taken from the
NASA/IPAC Extragalactic Database (NED).  Characteristics of the
gravitational lenses were checked with the CfA-Arizona Space
Telescope LEns Survey.  SIMBAD and HyperLeda databases was also used
to clarify the characteristics of some objects.  Also we simulated
and investigated a list of 4000 radio sources evenly distributed
around the celestial sphere.  We estimated the correlation matrices
between the vector spherical harmonics using the real as well as
modelled distribution of the radio sources.}
{A new list of physical characteristics of 4261 astrometric
radio sources, including all 717 ICRF-Ext.2 sources has been compiled.
Comparison of our data of optical characteristics with the official
International Earth Rotation and Reference Systems Service (IERS) list
showed significant discrepancies for about half of 667 common sources.
Finally, we found that asymmetry in the radio sources distribution
between hemispheres could cause significant correlation between the
vector spherical harmonics, especially if the case of sparse
distribution of the sources with high redshift.
We also identified radio sources having many-year observation history and lack redshift.
This sources should be urgently observed at large optical telescopes.}
{The list of optical characteristics created in this paper is recommended
for use as a supplement material for the next International Celestial Reference
Frame (ICRF) realization. It can be also effectively used for cosmological studies
and planning of observing programs both in radio and optics.}


\maketitle


\section{Introduction}

Information on physical characteristics of the astrometric radio sources
is important for planning of VLBI experiments and analysis
of VLBI data to do a research in cosmology, kinematics of the
Universe, etc.  In particular, the primary mainspring to this work
was a support of the investigation of the systematic effects in
apparent motion of the astrometric radio sources observed by VLBI
(Gwinn et al. 1997, MacMillan 2005, Titov 2008a, Titov 2008b).

The official list of the physical characteristics of the ICRF radio
sources is supported by the IERS (International Earth Rotation and Reference Systems Service)
ICRS (International Celestial Reference System) Product Center (Archinal et al. 1997).
The latest version of the IERS list is available in the
Internet\footnote{http://hpiers.obspm.fr/icrs-pc/info/car\_physique\_ext1}.
However this list has some deficiencies:
\begin{itemize}
\item Not all the sources observed in the framework of geodetic and
  astrometric experiments are included in the IERS list.
\item The characteristics of some sources in the IERS list
  are outdated or doubtful.
\end{itemize}

To overcome this problems, we performed a compilation of new list of the physical
characteristics of astrometric radio sources using the latest information.
Hereafter this list is referred to as OCARS (Optical Characteristics of Astrometric Radio Sources).

The list of radio sources with their positions was originally taken from the Goddard VLBI astrometric
catalogues\footnote{http://vlbi.gsfc.nasa.gov/solutions/astro}, version 2007c,
with addition of two absent ICRF-Ext.2 (Fey et al. 2004) sources 1039-474 and
1329-665 {(\bf hereafter we will use 8-char IERS designation HHMMsDDd which is
an abridged version of the IAU-compliant name 'IERS BHHMMsDDd')}.
In the last version, the source list was updated using the 2009a\_astro.cat catalogue
computed by Leonid Petrov\footnote{http://astrogeo.org/vlbi/solutions}.
It gives 4261 radio sources in total.

At this stage mainly the NASA/IPAC Extragalactic
Database\footnote{http://nedwww.ipac.caltech.edu/} (NED) was scoured.
Characteristics of the gravitational lenses were checked with the
CfA-Arizona Space Telescope LEns Survey\footnote{http://cfa-www.harvard.edu/glensdata/} (CASTLES).
Several sources were checked with SIMBAD\footnote{http://simbad.u-strasbg.fr/} the
HyperLeda\footnote{http://leda.univ-lyon1.fr/} databases.
In the OCARS list we have included only the optical characteristics of astrometric
radio sources: source type, redshift and visual magnitude.
The flux parameters are not included in our list because they are available from other
centers directly working on correlation and primary processing of the VLBI observations.

The OCARS was preliminary presented in (Malkin \& Titov 2008).
In this paper we investigated statistical properties of the list in more detail
and study their impact on the kinematic analysis of radio source motions.

Analysis of the radio source apparent motion revealed some
statistically significant systematics described by the vector
spherical harmonics of the first and second orders (dipole and
quadrupole effects, respectively) (Titov 2008a, Titov 2008b).
The dipole effect could be caused by the Galactocentric acceleration
of the Solar system (Gwinn et al. 1997; Sovers et al. 1998;
Kovalevsky 2003; Klioner 2003; Kopeikin \& Makarov 2006) or a
hypothetic acceleration of the Galaxy relative to the reference
quasars. The quadrupole harmonic, considered in details by
Kristian \& Sachs (1966), could be caused either of the primordial
gravitational waves or anisotropic expansion of the Universe.
This result was confirmed by Ellis et al. (1985) although they stated
``the major problem is that neither the distortion nor the proper motions
are likely to be measurable in practice in the foreseeable future''.
In this case the quadrupole effect should be redshift-dependent, and the
apparent proper motion will increase with redshift. However,
Pyne et al. (1996) and Gwinn et al. (1997) also discussed the gravitational
waves with the wavelength shorter than the Hubble length. Thus, the
proper motion, induced by the short-wavelength gravitational waves,
also might be constant over all redshifts.

Due to asymmetry of the astrometric radio source distribution
around the sky, the correlation between the vector
spherical harmonic components is not zero. Therefore, we studied the
effect of the asymmetry using the real uneven and simulated even
distribution of the sources.

The OCARS list can be used as a
supplement material for the second realization of the International
Celestial Reference Frame (ICRF2), as well as a database for
kinematic studies of the Universe and other related works, including
scheduling of dedicated IVS (International VLBI Service for Geodesy and Astrometry,
Schl\"ueter \& Behrend 2007) programs.

\section{Description of the OCARS}

Our primary interest is to get the redshift (z) for astrometric radio
sources to develop the previous studies
(Gwinn et al. 1997; MacMillan 2005; Titov 2008a, 2008b).
In those papers, redshift values were taken from the ICRF list
(Archinal et al. 1997).  However, as rather tiny effects in the source
motions are to be investigated, it is important to increase the
number of sources involved in the processing.  Searching the latest
astrophysical databases, primarily the NED, we could considerably
augment the list of astrometric radio sources with known redshift.
Nevertheless, more than half of the astrometric radio sources have
no determined redshift.

Evidently, the only direct way to get the redshift for other
most frequently observed astrometric sources is to organize a dedicated
observing program with large optical telescopes.  To help in
preparation of such a program, we also collect the source type and
its visual magnitude if this information is available.  Also, it
makes a sense to include in this observational program those sources
with existing but uncertain redshift values.

It should be noted, that not all astrometric radio sources were reliably
identified in the NED. We use the following procedure for identification.
In the first step, we search for sources by source name
using ``ICRF'' and ``IVS'' prefix.  So, we rely on the source
identification used in the literature and by the NED staff.  Then
about 500 sources, mostly from the VCS6 list, were searched by
position.  We take into account both the angular distance between the
VLBI and NED positions as well as their uncertainty in
the VLBI and NED positions. For some sources multiply NED objects within
the error level were found.  For 16 sources no appropriate
object was found in the NED, which is mentioned in the comments.
The problem of the source identification in the NED and
other astrophysical data bases hopefully will be solved after
official publication of the VCS6 catalogue.

The OCARS list is made available along with this paper in electronic form.

\section{Statistics}
\label{sect:stat}

The overall statistics of the OCARS is the following.

\vskip 2ex
\begin{center}
\noindent
\small
\tabcolsep=1pt
\begin{tabular}{lrr}
\underline{Number of sources:} && \\
total                               && 4261 (100\%~) \\
\quad N      & 2391 (56.1\%) \\
\quad S      & 1870 (43.9\%) \\
with known type                     && 2545 (59.7\%) \\
\quad AGN    & 1654 (65.0\%) & \\
\quad galaxy &  492 (19.3\%) & \\
\quad star   &   27 (~1.1\%) & \\
\quad other  &  372 (14.6\%) & \\
with known redshift                 && 1840 (43.2\%) \\
\quad $<=1$  &  853 (46.4\%) & \\
\quad $>1$   &  987 (53.6\%) & \\
\quad N      & 1195 (64.9\%) & \\
\quad S      &  645 (35.1\%) & \\
with known visual magnitude         && 2452 (57.5\%) \\
with known both z and magnitude     && 1789 (42.0\%) \\
with known z or magnitude           && 2503 (58.7\%) \\
with known magnitude and unknown z  && ~663 (15.6\%) \\
\end{tabular}
\end{center}
\vskip 2ex

\begin{figure}[ht!]
\centering
\resizebox{\hsize}{!}{\includegraphics[clip]{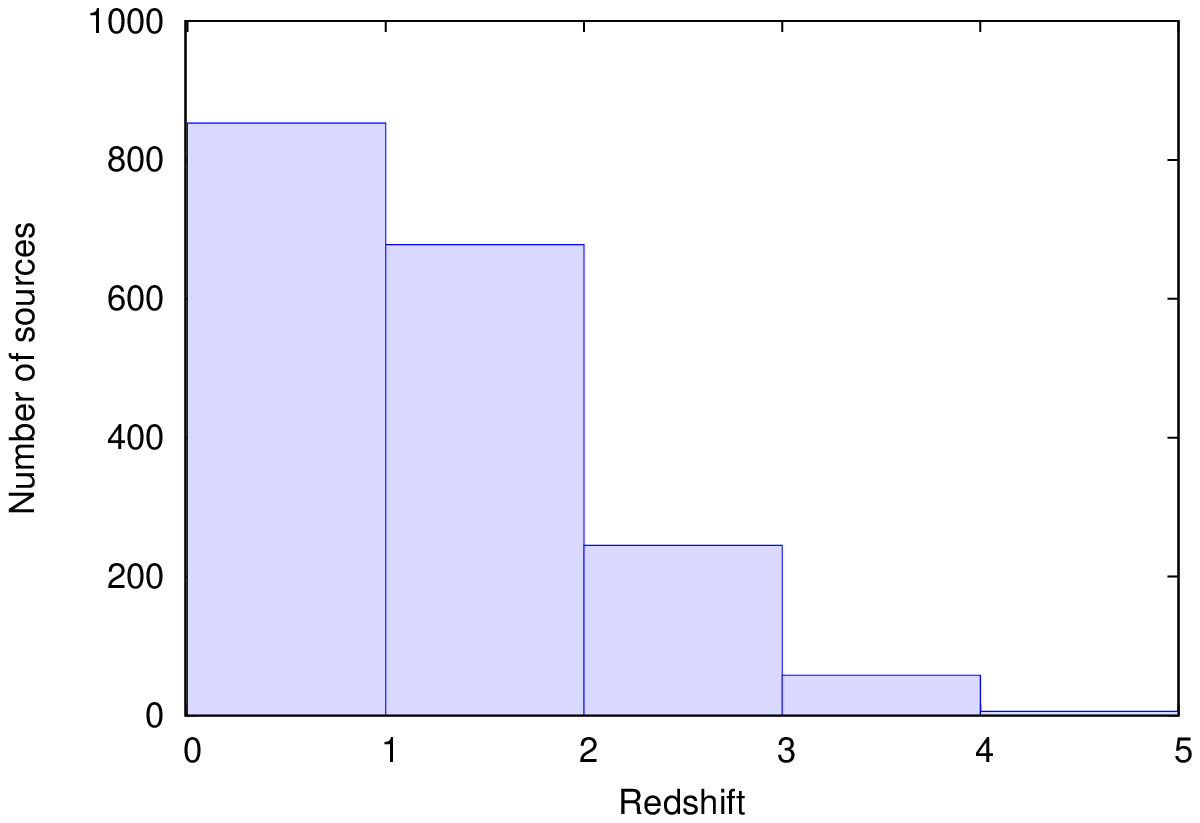}}
\hspace{5mm}
\resizebox{\hsize}{!}{\includegraphics[clip]{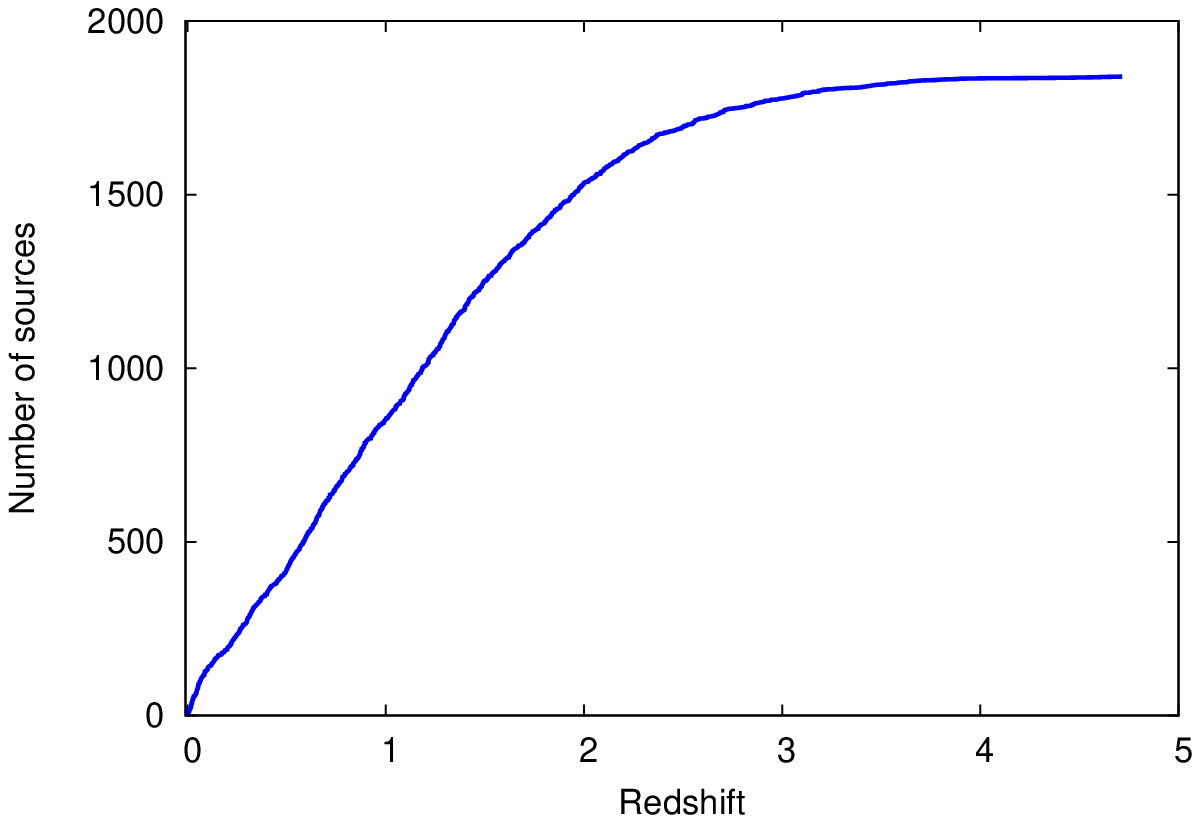}}
\caption{Distribution of the redshift ({\it top}) and cumulative
  number of sources ({\it bottom}).}
\label{fig:z_de_hist}
\end{figure}

\begin{figure}[ht!]
\centering
\resizebox{\hsize}{!}{\includegraphics[clip]{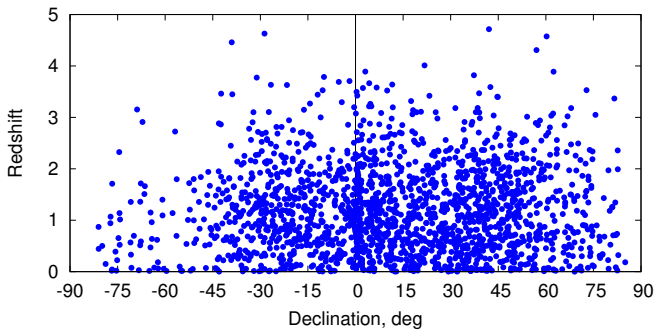}}
\caption{Distribution of the redshift over declination.}
\label{fig:z_de_cum}
\end{figure}

Figures \ref{fig:z_de_hist} and \ref{fig:z_de_cum} show the distribution
of the sources with known redshift.

\begin{figure}[ht!]
\centering
\resizebox{\hsize}{!}{\includegraphics[clip]{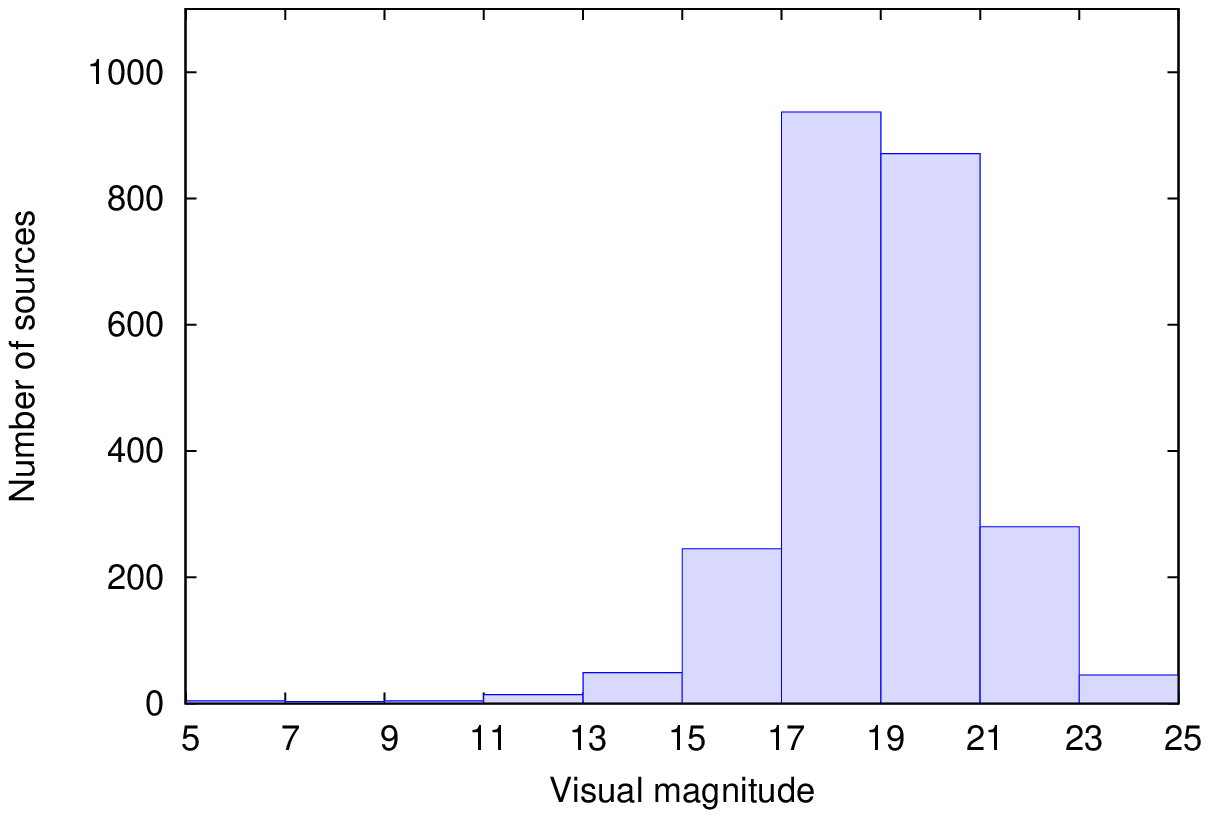}}
\hspace{5mm}
\resizebox{\hsize}{!}{\includegraphics[clip]{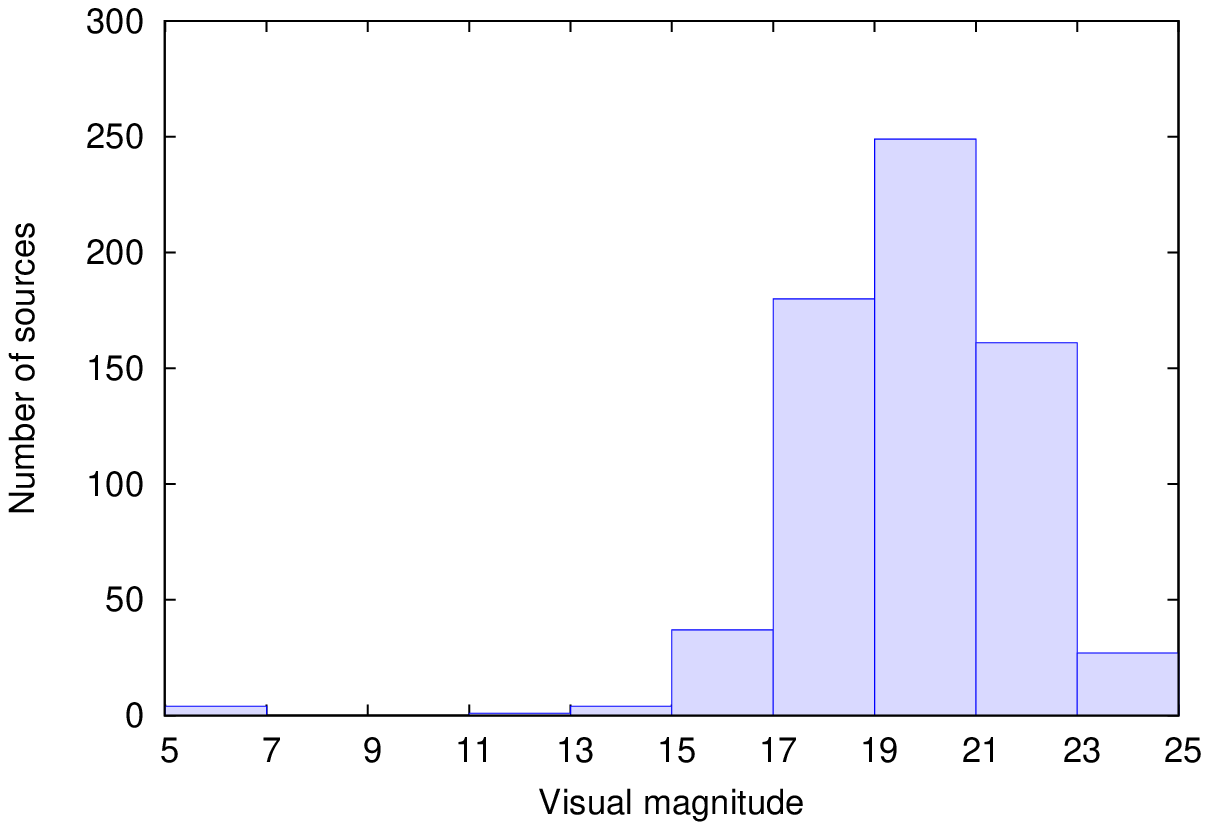}}
\caption{Distribution of the visual magnitude for all sources ({\it top})
and for sources without known redshift ({\it bottom}).}
\label{fig:v_hist}
\end{figure}

Figure \ref{fig:v_hist} shows the distribution of the visual magnitude.
The bottom part of the figure gives an impression about the magnitude
of the sources for which redshift is not yet determined.

Therefore, large observational projects for spectroscopy of the astrometric radio sources are very important.
Such a program is quite laborious taking into account a necessity of observations
of mostly rather weak sources and their careful identification (search for an
optical counterpart of radio sources).
So, it makes sense to create a list of radio sources which were intensively
observed during astrometric and geodetic VLBI programs, and lack of known redshift
to establish an order of priority for optical observations.
Such a list of high-priority sources is given in Table~\ref{tab:priority_list}.
It was compiled using the IVS observation statistics available at
\mbox{\tt http://www.gao.spb.ru/english/as/ac\_vlbi/}.
The list was sorted using the number of observations marked as good in the observational NGS cards
made during 24h sessions (in fact, in sessions of 18 hours and longer).
On the other hand, it seems to be reasonable to give a observation
priority to sources with reliably determined redshift, especially
in the Southern hemisphere, and sources already having a good observational history.

\begin{table}
\centering
\caption{High-priority list of radio sources lacking known redshift.}
\label{tab:priority_list}
\begin{tabular}{lc|lc|lc}
\hline
\hline
\multicolumn{1}{c}{Source} & Nobs & \multicolumn{1}{c}{Source} & Nobs &
\multicolumn{1}{c}{Source} & Nobs\\
\hline
1357+769 & 202005 & 0426+273 & 994 & 0017+200 & 356 \\
1300+580 &  75748 & 0019+058 & 941 & 0206+136 & 342 \\
0749+540 &  67718 & 1706-174 & 919 & 2051+745 & 337 \\
0718+792 &  35867 & 1922-224 & 908 & 1651+391 & 326 \\
0300+470 &  25613 & 1746+470 & 850 & 0459+135 & 323 \\
1923+210 &  25324 & 2337+264 & 723 & 1647-296 & 313 \\
0735+178 &  16356 & 0134+311 & 708 & 1243-160 & 310 \\
0556+238 &  12553 & 0826-373 & 693 & 0729+259 & 310 \\
1057-797 &   6623 & 0554+242 & 635 & 1036+054 & 310 \\
0656+082 &   5278 & 0818-128 & 633 & 1951+355 & 304 \\
2229+695 &   4759 & 0458+138 & 617 & 0332+078 & 303 \\
0657+172 &   4667 & 0524+034 & 592 & 0822+137 & 301 \\
0454+844 &   4296 & 1754+155 & 567 & 0602+405 & 300 \\
1657-261 &   2735 & 1823+689 & 536 & 1909+161 & 296 \\
1221+809 &   2258 & 1926+087 & 531 & 1424+240 & 293 \\
1929+226 &   2087 & 0759+183 & 530 & 1734+508 & 291 \\
2000+472 &   2086 & 1656-075 & 517 & 0437-454 & 281 \\
0430+289 &   2047 & 1639-062 & 516 & 1251-713 & 277 \\
0648-165 &   1973 & 0055-059 & 509 & 0743+277 & 268 \\
0109+224 &   1954 & 1604-333 & 502 & 0601+245 & 267 \\
1751+288 &   1906 & 1443-162 & 497 & 0259+121 & 263 \\
0420+417 &   1893 & 1817-254 & 490 & 0951+268 & 260 \\
2150+173 &   1789 & 1327+321 & 488 & 0723+219 & 250 \\
1815-553 &   1789 & 2013+163 & 479 & 2302+232 & 248 \\
1459+480 &   1782 & 0341+158 & 472 & 0613+570 & 242 \\
0302+625 &   1660 & 0415+398 & 470 & 0302-623 & 232 \\
0544+273 &   1634 & 0733-174 & 468 & 0700-197 & 226 \\
0920+390 &   1575 & 0854-108 & 453 & 1433+304 & 225 \\
0159+723 &   1508 & 1602-115 & 447 & 1830+139 & 222 \\
0440+345 &   1467 & 0854+213 & 446 & 1705+135 & 215 \\
1738+476 &   1413 & 1922+155 & 438 & 0629+160 & 215 \\
2021+317 &   1335 & 1947+079 & 435 & 1013+127 & 215 \\
1307+121 &   1296 & 2000+148 & 433 & 0728+249 & 214 \\
1213-172 &   1285 & 0115-214 & 424 & 0025+197 & 213 \\
0308-611 &   1280 & 1101-536 & 405 & 1344+188 & 205 \\
0722+145 &   1269 & 1409+218 & 400 & 1451+270 & 200 \\
0506+101 &   1233 & 0912+297 & 397 & 0516-621 & 200 \\
1932+204 &   1128 & 0627-199 & 376 & 1156-094 & 198 \\
0039+230 &   1123 & 0530-727 & 375 & 1412-368 & 196 \\
2030+547 &   1056 & 0725+219 & 369 & 0627+176 & 195 \\
\hline
\end{tabular}
\end{table}

\section{Comparison with the IERS list}

We have compared the OCARS with the IERS list and found 667 common radio sources.
All the sources are present in the IERS list.
Comparison of these two lists results in rather large discrepancy.
\begin{itemize}
\item The first evident difference is in the number of radio sources.
The OCARS list contains 40 extra ICRF sources plus several hundreds other
sources, 4261 vs. 667 objects in total, 2503 vs. 555 objects with
known redshift or visual magnitude.
\item Unlike the IERS list, we did not try to trace all the details of the
Active Galactic Nuclei (AGN) classification that are not always
stable and unambiguous.  So, all the quasars and BL object are
designated as AGN.
\item Redshift for 55 more ICRF sources were found; redshift
for 4 sources presented in the IERS list were not included in our list
for various reasons; for 30 sources redshift differs more than
by 0.01; the largest differences are 1.26 (1903-802), 1.20
(1600+431), 0.70 (0646-306).
\item Visual magnitudes for 70 more ICRF sources were found;
magnitudes for 2 sources were not confirmed in our list; for 195
sources magnitude differs more than by 0.5; the largest
differences are 5.2 (1758-651), 5.0 (1156-094, 1322-427), 3.9 (0241+622).
\end{itemize}

Further investigation has to be made to clarify all found
discrepancies between two lists.  It should be mentioned that we
consider as important and useful for a user to provide a detailed
comment in case of doubtful or ambiguous published data.

\section{Comparison with the LQAC}

We also performed a comparison of the OCARS list with recently published
Large Quasar Astrometric Catalogue (LQAC) (Souchay et al. 2009).
This catalogue contains information on redshift and luminosity
at optical and radio wavelengths for 113666 quasars.

First, it should be noticed that more than one third of the astrometric sources
are not quasars (see section~\ref{sect:stat}, hence not all of them are contained in the LQAC.
Nevertheless, this comparison seems to be mutually interesting because the information sources
used to compile both lists evidently have a large intersection, but are not fully the same.
Also, such a check allows us to reveal possible mistakes, overlooked literature, etc.

Since LQAC has no source names, the source identification was made using their position in both
catalogues, which corresponds to the method used by the authors of LQAC.
Table~\ref{tab:OCARS-QLAC} show a summary of comparison results for five 5 different search radii.
Here we compared only redshift values as the most important parameter.
The redshift difference of 0.01 was used to detect the significant redshift difference between
two compared catalogues.

\begin{table}
\caption{Statistics of comparison of OCARS (1) with LQAC (2).}
\label{tab:OCARS-QLAC}
\tabcolsep=4pt
\begin{tabular}{lccccc}
\hline
\hline
\multicolumn{1}{l}{Number of sources ...} & \multicolumn{5}{c}{Search radius, arcsec} \\
\cline{2-6}
                                &   1  &   2  &   5  &  10  &   30 \\
\hline
not found in (2)                &  669 &  650 &  641 &  635 &  631 \\
without z in both catalogues    & 1809 & 1811 & 1814 & 1817 & 1821 \\
with z in (1) only              &  547 &  548 &  532 &  526 &  520 \\
with z in (2) only              &    8 &    8 &    8 &   10 &   10 \\
with different z in (1) and (2) &   44 &   48 &   47 &   47 &   50 \\

\hline
\end{tabular}
\end{table}

The most interesting result to us are sources having redshift in the LQAC and lack it in the OCARS.
Those sources are
0003-302,  0118-283,  1026-084,  1202+527,  1225+028,  1332+031,  1422+231, 1555+030, 1639-062, and 2256-084.
The first analysis showed that the sources 0003-302 and 1026-084 are located in very populated sky region,
and maybe more careful identification is needed.
The redshift value for the former is referred to a private communication.
The source 1422+231 is a component of a gravitational lens according to CASTLES.
These and other discrepancies found during this comparison will be analyzed during preparation of the next OCARS version.

An interesting point is the number of the OCARS objects not found in the LQAC.
One can see that about 85\% OCARS objects were found in the LQAC, which is
much greater than the number of OCARS object with known type.
On the other hand, the number of the OCARS objects not found in the LQAC
is less than the number of OCARS non-AGN objects.
It may be worth re-visiting the object classification in both catalogues, in fact in LQAC and NED.
Such a large work is out of the scope of this study however.

\section{Vector spherical functions}

The paucity of the radio sources with declination less than $-30^\circ$
(Fig~\ref{fig:z_de_cum}) may result in difficulties for astrometric
analysis of radio catalogues.
A worse precision of the radio source coordinates in the Southern
hemisphere has been reported in papers on
the International Celestial Reference Frame (Ma et al. 1998).
It may cause more dramatic impact on the analysis of the
apparent proper motion of the reference radio sources.

The ICRF consists of a set of highly
accurate positions of reference radio sources as determined by the
VLBI techniques.  The observed radio sources are very distant,
therefore when the ICRF is made, their physical proper motion is
assumed to be negligible (less then 1 $\mu$as/year) and their
positions are practically stable. However, the motion of relativistic
jets from the active extragalactic nuclei can mimic the proper motion
of the observed radio sources (with a magnitude up to 0.5 mas/yr)
(see e.g. Marcaide et al. 1985; Alberdi et al. 1993; Fey et al. 1997,
Feissel-Vernier 2003; Titov 2007; MacMillan \& Ma 2007).
Moreover, some tiny systematic effects (dipole and quadrupole) in the
apparent proper motion have been observed (Titov 2008a, 2008b). A
study of these apparent proper motion would become an important part
of the future fundamental reference frame making up.

The dipole and quadrupole systematic effect in the radio source
apparent motion were studied previously (Gwinn et al. 1997; MacMillan 2005;
Titov 2008a, 2008b) using the expansion of the apparent motion
field on vector spherical harmonics. Unfortunately, due to uneven
distribution of the reference radio sources over the sky the results
could be corrupted by the correlation between estimated parameters.
However, the correlation matrices have not been studied carefully so far.

The vector spherical harmonics Hill (1954) are used to study the
systematic effect in a proper motion of celestial objects (see e.g.
Mignard \& Morando 1990; Gwinn et al. 1997; Vityazev \& Shuksto 2004;
Vityazev \& Tsvetkov 2009).

Let us consider $\vec{F}(\alpha,\delta)$ as a vector field of a
sphere described by the components of the apparent proper motion
vector ($\mu_{\alpha}\cos\delta$, $\mu_{\delta}$)

\begin{equation}
  \vec{F}({\alpha},{\delta}) =
   {\mu_{\alpha}}\cos{\delta}\vec{e}_{\alpha} + {\mu_{\delta}}\vec{e}_{\delta} \,,
\label{eq:F1}
\end{equation}
where $\vec{e}_{\alpha}$, $\vec{e}_{\delta}$ --- unit vectors.
A vector field of spherical functions $\vec{F}$($\alpha$,$\delta$)
can be approximated by the vector spherical functions as follows
\begin{equation}
\vec{F}({\alpha},{\delta}) = \sum\limits_{l=1}^{\infty}
  \sum\limits_{m=-l}^{l} (a_{l,m}^{E}\vec{Y}_{l,m}^{E} + a_{l,m}^{M}\vec{Y}_{l,m}^{M}) \,,
\label{eq:F}
\end{equation}
where $\vec{Y}_{l,m}^{E}$, $\vec{Y}_{l,m}^{M}$ --- the `electric' and
`magnetic' transverse vector spherical functions, respectively:
\begin{equation}
\begin{array}{rcl}
\vec{Y}_{l,m}^{E} &=& \frac{1}{\displaystyle\sqrt{l(l+1)}} \left(\displaystyle\frac{\partial{V_{l,m}
  ({\alpha},{\delta})}}{\partial{\alpha}\cos{\delta}}\vec{e}_{\alpha}  + \frac{\partial{V_{l,m}({\alpha},{\delta})}}
  {\partial{\delta}}\vec{e}_{\delta} \right) \,, \\[0.5em]

\vec{Y}_{l,m}^{M} &=& \frac{1}{\displaystyle\sqrt{l(l+1)}} \left(\displaystyle\frac{\partial{V_{l,m}
  ({\alpha},{\delta})}}{\partial{\delta}}\vec{e}_{\alpha} - \frac{\partial{V_{l,m}({\alpha},{\delta})}}
  {\partial{\alpha}\cos{\delta}}\vec{e}_{\delta} \right) \,.
\end{array}
\label{eq:Y}
\end{equation}

The function $V_{l,m}$($\alpha$,$\delta)$ is given by
\begin{equation}
V_{l,m}({\alpha},{\delta}) = (-1)^{m}\sqrt{\frac{(2l+1)(l-m)!}
{4{\pi}(l+m)!}}P_{l}^{m}(\sin{\delta}){\exp(im{\alpha}}) \,,
\label{eq:V}
\end{equation}
where $P_{l}^{m}(sin\delta)$ --- the associated Legendre functions

The coefficients of expansion $a_{l,m}^{E}$, $a_{l,m}^{M}$ to be
estimated as follows
\begin{equation}
\begin{array}{rcl}
a_{l,m}^{E} &=&
  \int_{0}^{2{\pi}}\int_{-\frac{\pi}{2}}^{\frac{\pi}{2}}\vec{F}({\alpha},{\delta})
  \vec{Y}_{l,m}^{E{\ast}}({\alpha},{\delta})\cos{\delta}\, d{\alpha}d{\delta} \,, \\[0.5em]
a_{l,m}^{M} &=&
  \int_{0}^{2{\pi}}\int_{-\frac{\pi}{2}}^{\frac{\pi}{2}}\vec{F}({\alpha},{\delta})
  \vec{Y}_{l,m}^{M{\ast}}({\alpha},{\delta})\cos{\delta} \, d{\alpha}d{\delta} \,,
\end{array}
\label{eq:a}
\end{equation}
where $\ast$ means a complex conjugation. This system of equations
can be solved by the least squares method. In this research the
coefficients are estimated as global parameters from a large set of
VLBI data.

The three `electric' spherical harmonics for l=1 produces the dipole
effect corresponding to the acceleration of the Solar system. The
three l=1 `magnetic' harmonics describe a rotation of the set of
quasars.  This rotation is not separable from the Earth's rotation,
and, therefore, not estimated. The five l=2 `electric' spherical
harmonics correspond either to the gravitational waves or to the
Universe anisotropic expansion, and, finally, the five l=2 `magnetic'
spherical harmonics correspond only to the gravitational waves. In
total, 13 components are to be estimated.

\begin{figure*}
\centering
{\tiny
\parbox[b]{0.98\columnwidth}{
\raisebox{8em}{\includegraphics[clip,width=0.52\columnwidth,angle=-90]{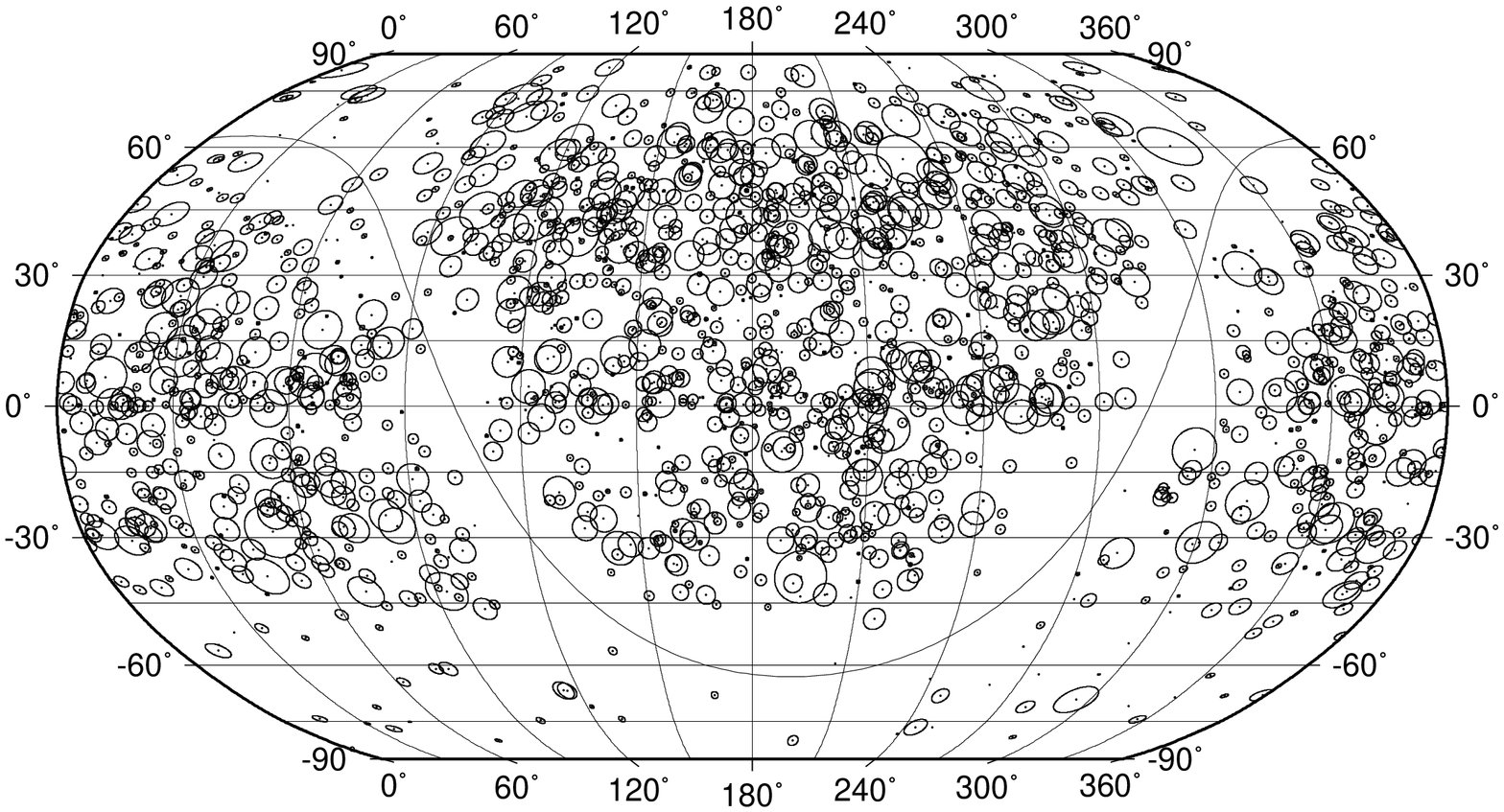}}
}
\hspace{0.03\columnwidth}
\parbox[t]{1.006\columnwidth}{
\tabcolsep=1pt
\begin{tabular}{r|rrrrrrrrrrrrr}
& $a_{1, 1}^{E}$ & $a_{1,-1}^{E}$ & $a_{1, 0}^{E}$ & $a_{2, 1}^{E}$ & $a_{2,-1}^{E}$ &
  $a_{2, 2}^{E}$ & $a_{2,-2}^{E}$ & $a_{2, 0}^{E}$ & $a_{2, 1}^{M}$ & $a_{2,-1}^{M}$ &
  $a_{2, 2}^{M}$ & $a_{2,-2}^{M}$ & $a_{2, 0}^{M}$ \\
\hline
$a_{1, 1}^{E}$ & 1.00 &      &      &      &      &      &      &      &      &      &      &      &      \\
$a_{1,-1}^{E}$ & 0.02 & 1.00 &      &      &      &      &      &      &      &      &      &      &      \\
$a_{1, 0}^{E}$ &-0.10 &-0.02 & 1.00 &      &      &      &      &      &      &      &      &      &      \\
$a_{2, 1}^{E}$ & 0.02 &-0.26 &-0.02 & 1.00 &      &      &      &      &      &      &      &      &      \\
$a_{2,-1}^{E}$ &-0.24 & 0.01 & 0.01 &-0.05 & 1.00 &      &      &      &      &      &      &      &      \\
$a_{2, 2}^{E}$ &-0.03 & 0.12 &-0.02 &-0.03 & 0.05 & 1.00 &      &      &      &      &      &      &      \\
$a_{2,-2}^{E}$ & 0.08 &-0.04 &-0.05 & 0.04 & 0.08 &-0.03 & 1.00 &      &      &      &      &      &      \\
$a_{2, 0}^{E}$ &-0.08 &-0.01 &-0.26 & 0.01 & 0.03 & 0.02 & 0.10 & 1.00 &      &      &      &      &      \\
$a_{2, 1}^{M}$ & 0.02 &-0.04 &-0.07 & 0.01 &-0.11 & 0.02 & 0.03 &-0.12 & 1.00 &      &      &      &      \\
$a_{2,-1}^{M}$ & 0.05 &-0.28 & 0.01 & 0.17 &-0.02 &-0.12 & 0.03 & 0.02 &-0.04 & 1.00 &      &      &      \\
$a_{2, 2}^{M}$ & 0.13 & 0.07 &-0.23 & 0.02 & 0.05 & 0.01 & 0.03 & 0.06 &-0.03 &-0.03 & 1.00 &      &      \\
$a_{2,-2}^{M}$ &-0.09 & 0.07 & 0.11 &-0.06 &-0.01 & 0.02 &-0.02 &-0.03 &-0.01 &-0.08 &-0.06 & 1.00 &      \\
$a_{2, 0}^{M}$ &-0.03 & 0.16 & 0.00 & 0.09 &-0.02 & 0.03 &-0.02 & 0.00 &-0.01 &-0.05 & 0.03 & 0.11 & 1.00 \\
\end{tabular}}
\vskip 2em
\parbox[b]{0.98\columnwidth}{
\raisebox{8em}{\includegraphics[clip,width=0.52\columnwidth,angle=-90]{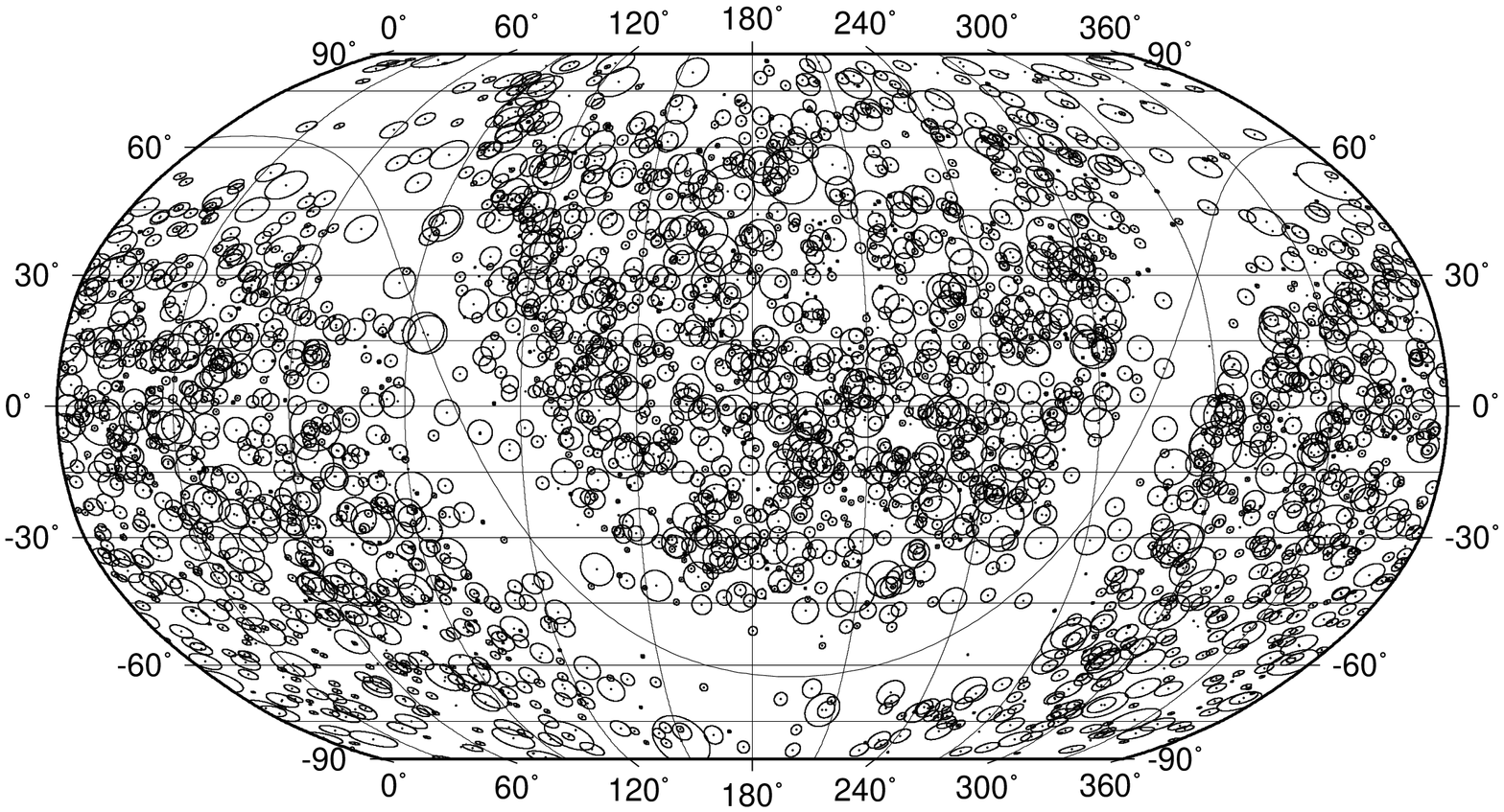}}
}
\hspace{0.03\columnwidth}
\parbox[t]{1.006\columnwidth}{
\tabcolsep=1pt
\begin{tabular}{r|rrrrrrrrrrrrr}
& $a_{1, 1}^{E}$ & $a_{1,-1}^{E}$ & $a_{1, 0}^{E}$ & $a_{2, 1}^{E}$ & $a_{2,-1}^{E}$ &
  $a_{2, 2}^{E}$ & $a_{2,-2}^{E}$ & $a_{2, 0}^{E}$ & $a_{2, 1}^{M}$ & $a_{2,-1}^{M}$ &
  $a_{2, 2}^{M}$ & $a_{2,-2}^{M}$ & $a_{2, 0}^{M}$ \\
\hline
$a_{1, 1}^{E}$ & 1.00 &      &      &      &      &      &      &      &      &      &      &      &      \\
$a_{1,-1}^{E}$ & 0.03 & 1.00 &      &      &      &      &      &      &      &      &      &      &      \\
$a_{1, 0}^{E}$ &-0.09 &-0.02 & 1.00 &      &      &      &      &      &      &      &      &      &      \\
$a_{2, 1}^{E}$ &-0.02 & 0.02 & 0.04 & 1.00 &      &      &      &      &      &      &      &      &      \\
$a_{2,-1}^{E}$ & 0.02 &-0.01 &-0.01 &-0.04 & 1.00 &      &      &      &      &      &      &      &      \\
$a_{2, 2}^{E}$ & 0.01 &-0.01 &-0.02 & 0.02 & 0.06 & 1.00 &      &      &      &      &      &      &      \\
$a_{2,-2}^{E}$ & 0.00 &-0.01 & 0.00 & 0.04 & 0.11 &-0.02 & 1.00 &      &      &      &      &      &      \\
$a_{2, 0}^{E}$ & 0.00 & 0.01 & 0.00 &-0.01 &-0.04 & 0.01 & 0.04 & 1.00 &      &      &      &      &      \\
$a_{2, 1}^{M}$ &-0.06 &-0.04 &-0.09 & 0.00 & 0.02 &-0.02 &-0.01 & 0.00 & 1.00 &      &      &      &      \\
$a_{2,-1}^{M}$ & 0.04 &-0.12 & 0.01 &-0.02 & 0.00 &-0.01 &-0.02 &-0.01 &-0.04 & 1.00 &      &      &      \\
$a_{2, 2}^{M}$ & 0.14 & 0.08 &-0.16 &-0.02 &-0.01 & 0.00 &-0.01 & 0.00 &-0.01 &-0.01 & 1.00 &      &      \\
$a_{2,-2}^{M}$ &-0.09 & 0.04 & 0.07 & 0.00 &-0.02 &-0.01 &-0.01 &-0.02 &-0.01 &-0.05 &-0.04 & 1.00 &      \\
$a_{2, 0}^{M}$ &-0.02 & 0.16 & 0.00 & 0.00 & 0.00 & 0.00 &-0.02 & 0.00 &-0.02 &-0.06 & 0.02 & 0.05 & 1.00 \\
\end{tabular}}
\vskip 2em
\parbox[b]{0.98\columnwidth}{
\raisebox{8em}{\includegraphics[clip,width=0.52\columnwidth,angle=-90]{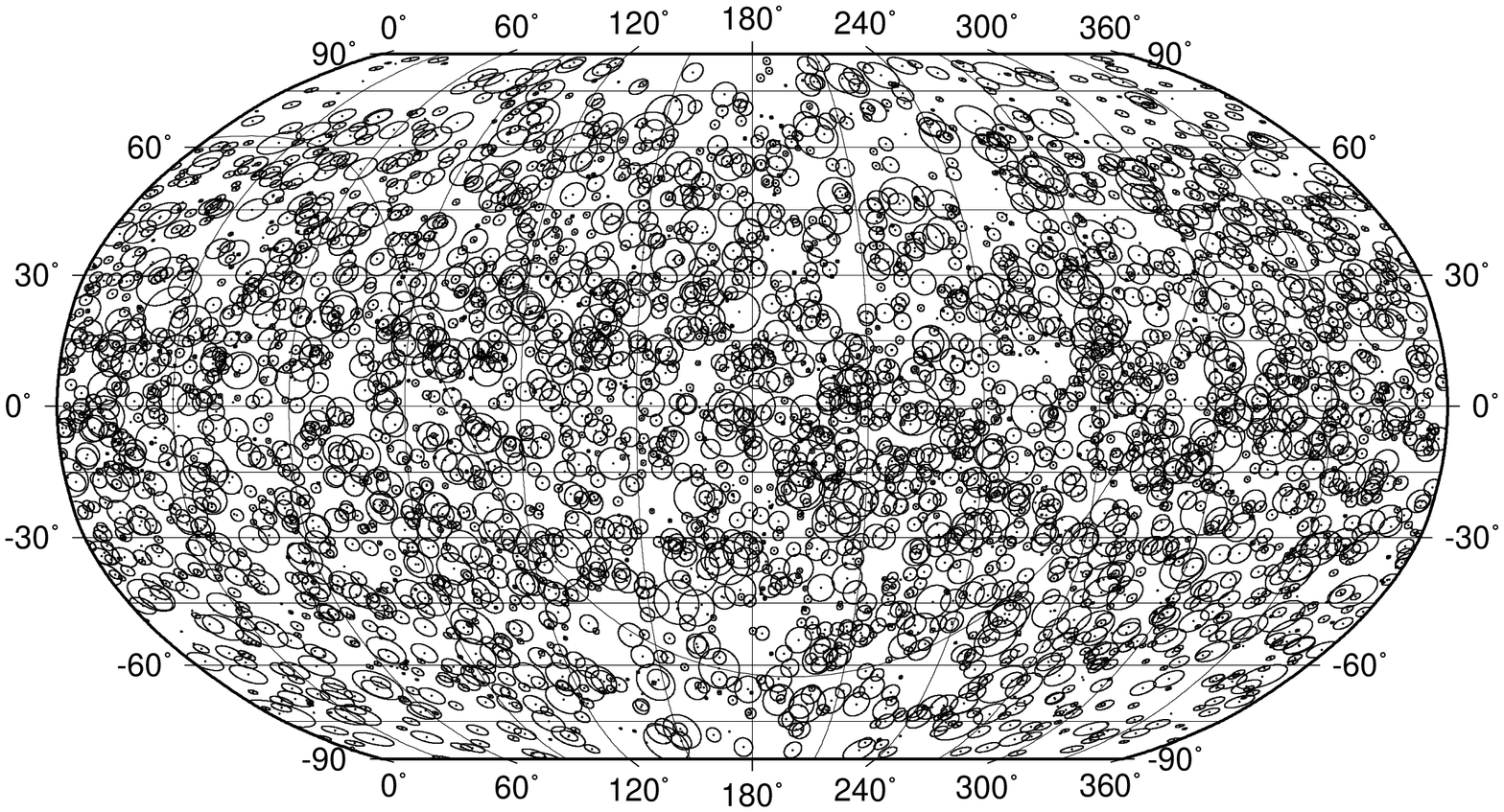}}
}
\hspace{0.03\columnwidth}
\parbox[t]{1.006\columnwidth}{
\tabcolsep=1pt
\begin{tabular}{r|rrrrrrrrrrrrr}
& $a_{1, 1}^{E}$ & $a_{1,-1}^{E}$ & $a_{1, 0}^{E}$ & $a_{2, 1}^{E}$ & $a_{2,-1}^{E}$ &
  $a_{2, 2}^{E}$ & $a_{2,-2}^{E}$ & $a_{2, 0}^{E}$ & $a_{2, 1}^{M}$ & $a_{2,-1}^{M}$ &
  $a_{2, 2}^{M}$ & $a_{2,-2}^{M}$ & $a_{2, 0}^{M}$ \\
\hline
$a_{1, 1}^{E}$ & 1.00 &      &      &      &      &      &      &      &      &      &      &      &      \\
$a_{1,-1}^{E}$ & 0.00 & 1.00 &      &      &      &      &      &      &      &      &      &      &      \\
$a_{1, 0}^{E}$ & 0.00 & 0.01 & 1.00 &      &      &      &      &      &      &      &      &      &      \\
$a_{2, 1}^{E}$ &-0.01 & 0.00 & 0.02 & 1.00 &      &      &      &      &      &      &      &      &      \\
$a_{2,-1}^{E}$ & 0.01 &-0.01 &-0.01 & 0.00 & 1.00 &      &      &      &      &      &      &      &      \\
$a_{2, 2}^{E}$ & 0.01 &-0.02 &-0.02 & 0.00 & 0.01 & 1.00 &      &      &      &      &      &      &      \\
$a_{2,-2}^{E}$ &-0.01 &-0.01 & 0.01 & 0.01 & 0.00 & 0.01 & 1.00 &      &      &      &      &      &      \\
$a_{2, 0}^{E}$ & 0.01 & 0.00 & 0.00 &-0.01 &-0.01 &-0.01 &-0.02 & 1.00 &      &      &      &      &      \\
$a_{2, 1}^{M}$ &-0.01 & 0.00 & 0.01 & 0.00 & 0.01 &-0.01 &-0.02 & 0.02 & 1.00 &      &      &      &      \\
$a_{2,-1}^{M}$ & 0.00 & 0.01 &-0.01 &-0.01 & 0.00 & 0.00 &-0.02 & 0.00 & 0.00 & 1.00 &      &      &      \\
$a_{2, 2}^{M}$ & 0.00 & 0.02 & 0.00 &-0.02 &-0.02 &-0.01 &-0.01 &-0.01 & 0.01 & 0.01 & 1.00 &      &      \\
$a_{2,-2}^{M}$ &-0.01 & 0.00 & 0.00 & 0.01 &-0.01 &-0.01 & 0.01 &-0.02 &-0.01 & 0.01 & 0.01 & 1.00 &      \\
$a_{2, 0}^{M}$ & 0.03 &-0.01 & 0.00 &-0.02 & 0.00 &-0.01 &-0.02 & 0.00 &-0.01 &-0.01 &-0.01 &-0.02 & 1.00 \\
\end{tabular}}
}
\caption{Sky coverage and correlation matrix for 13 vector spherical harmonics of first and second
order for real source distribution of the 1809 radio sources with known redshift for real
distribution ({\it top}) and two modelled distributions of 3123 radio source with the avoidance zone ({\it middle})
and of 4000 radio sources without the avoidance zone ({\it bottom}).
The circle size corresponds to the redshift value.}
\label{fig:distr-corr_all}
\end{figure*}

We studied both real and modelled distributions of the radio sources
to calculate the correlations between the spherical harmonics (see
Fig~\ref{fig:distr-corr_all}).  For modelling, we initially created a
set of 4000 radio sources uniformly distributed over the sky.  Then
we assigned redshift to all simulated objects in a such way that the
redshift values have the same distribution as real ones.  Another
model set of sources was created from the first one by thinning of
the sources within $\pm15^\circ$ zone along the Galactic equator, to
get an `avoidance zone' similar to the real distribution.
These distributions are shown in Fig~\ref{fig:distr-corr_all}.

\begin{table}
\centering
\caption{Statistic of the model and real distributions of the radio
sources over four redshift intervals. The number of radio sources and
the maximum correlation between the thirteen estimated parameters are shown.}
\begin{tabular}{ccccc}
\hline
\hline
Zone & \multicolumn{2}{c}{Model} & \multicolumn{2}{c}{Real} \\
     &  N  & Max corr & N &   Max corr \\
\hline
$0<z<1$ & 1431 & 0.15 &   835  &  0.27 \\
$1<z<2$ & 1183 & 0.15 &   671  &  0.33 \\
$2<z<3$ & ~398 & 0.18 &   240  &  0.34 \\
$3<z<4$ & ~100 & 0.19 &   ~57  &  0.51 \\
\hline
\end{tabular}
\end{table}

\begin{figure*}
\centering
{\tiny
\parbox[b]{0.98\columnwidth}{
\raisebox{8em}{\includegraphics[clip,width=0.52\columnwidth,angle=-90]{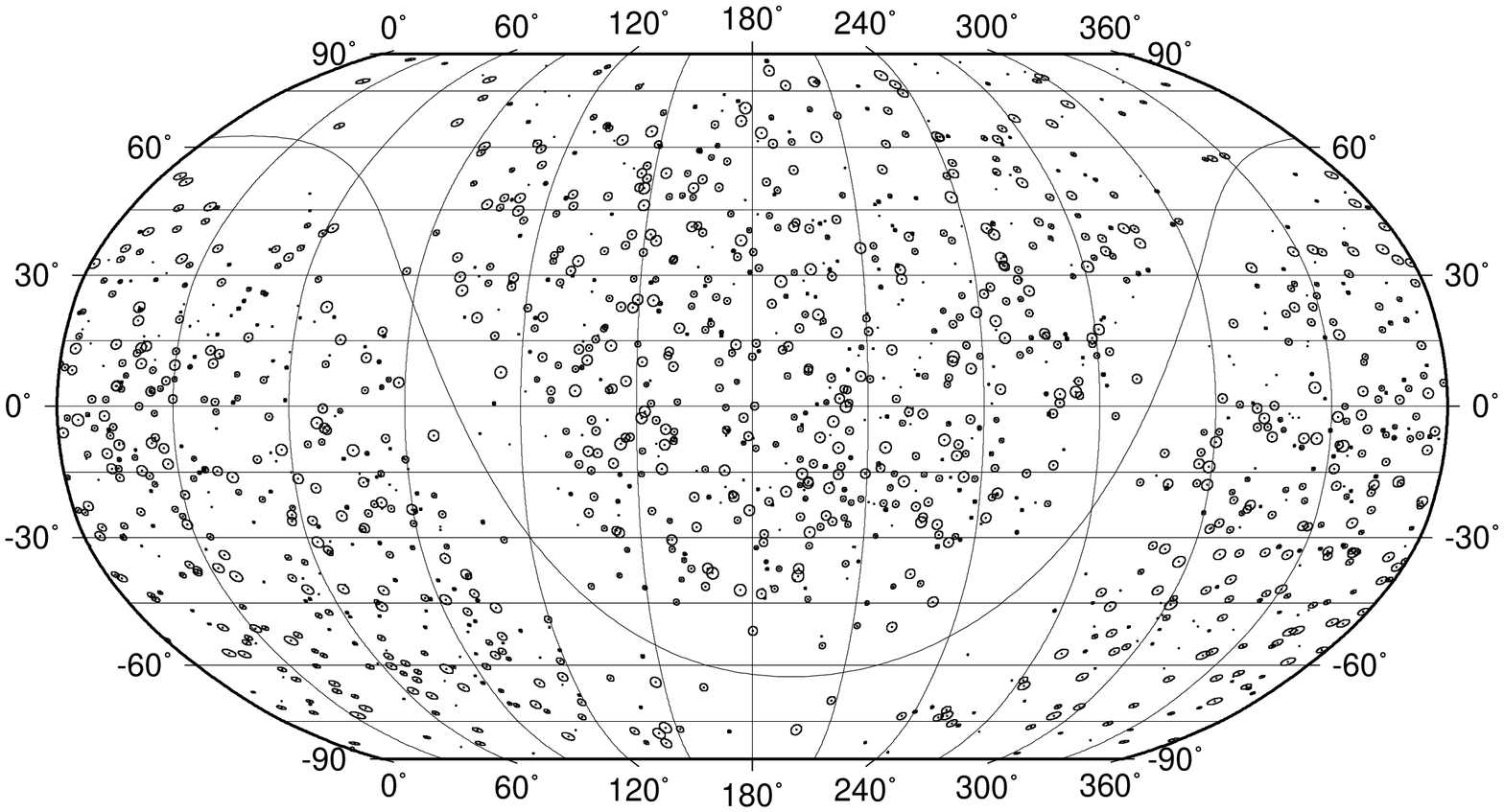}}
}
\hspace{0.03\columnwidth}
\parbox[t]{1.006\columnwidth}{
\tabcolsep=1pt
\begin{tabular}{r|rrrrrrrrrrrrr}
& $a_{1, 1}^{E}$ & $a_{1,-1}^{E}$ & $a_{1, 0}^{E}$ & $a_{2, 1}^{E}$ & $a_{2,-1}^{E}$ &
  $a_{2, 2}^{E}$ & $a_{2,-2}^{E}$ & $a_{2, 0}^{E}$ & $a_{2, 1}^{M}$ & $a_{2,-1}^{M}$ &
  $a_{2, 2}^{M}$ & $a_{2,-2}^{M}$ & $a_{2, 0}^{M}$ \\
\hline
$a_{1, 1}^{E}$ & 1.00 &      &      &      &      &      &      &      &      &      &      &      &      \\
$a_{1,-1}^{E}$ & 0.03 & 1.00 &      &      &      &      &      &      &      &      &      &      &      \\
$a_{1, 0}^{E}$ &-0.10 &-0.04 & 1.00 &      &      &      &      &      &      &      &      &      &      \\
$a_{2, 1}^{E}$ &-0.01 & 0.04 & 0.03 & 1.00 &      &      &      &      &      &      &      &      &      \\
$a_{2,-1}^{E}$ & 0.03 &-0.01 &-0.01 &-0.04 & 1.00 &      &      &      &      &      &      &      &      \\
$a_{2, 2}^{E}$ &-0.01 &-0.01 &-0.01 & 0.03 & 0.08 & 1.00 &      &      &      &      &      &      &      \\
$a_{2,-2}^{E}$ & 0.02 & 0.00 & 0.00 & 0.04 & 0.10 & 0.00 & 1.00 &      &      &      &      &      &      \\
$a_{2, 0}^{E}$ & 0.00 & 0.02 & 0.03 &-0.01 &-0.03 & 0.00 & 0.07 & 1.00 &      &      &      &      &      \\
$a_{2, 1}^{M}$ &-0.07 &-0.04 &-0.09 & 0.00 & 0.01 & 0.00 &-0.02 &-0.01 & 1.00 &      &      &      &      \\
$a_{2,-1}^{M}$ & 0.04 &-0.15 & 0.04 &-0.02 & 0.00 &-0.02 &-0.04 &-0.02 &-0.04 & 1.00 &      &      &      \\
$a_{2, 2}^{M}$ & 0.13 & 0.09 &-0.19 & 0.00 &-0.01 & 0.00 & 0.00 & 0.00 &-0.02 &-0.03 & 1.00 &      &      \\
$a_{2,-2}^{M}$ &-0.12 & 0.03 & 0.07 &-0.01 &-0.01 & 0.00 &-0.01 &-0.01 & 0.00 &-0.05 &-0.03 & 1.00 &      \\
$a_{2, 0}^{M}$ &-0.06 & 0.17 & 0.00 & 0.01 & 0.02 & 0.01 &-0.01 & 0.00 &-0.02 &-0.06 & 0.01 & 0.09 & 1.00 \\
\end{tabular}}
\vskip 2em
\parbox[b]{0.98\columnwidth}{
\raisebox{8em}{\includegraphics[clip,width=0.52\columnwidth,angle=-90]{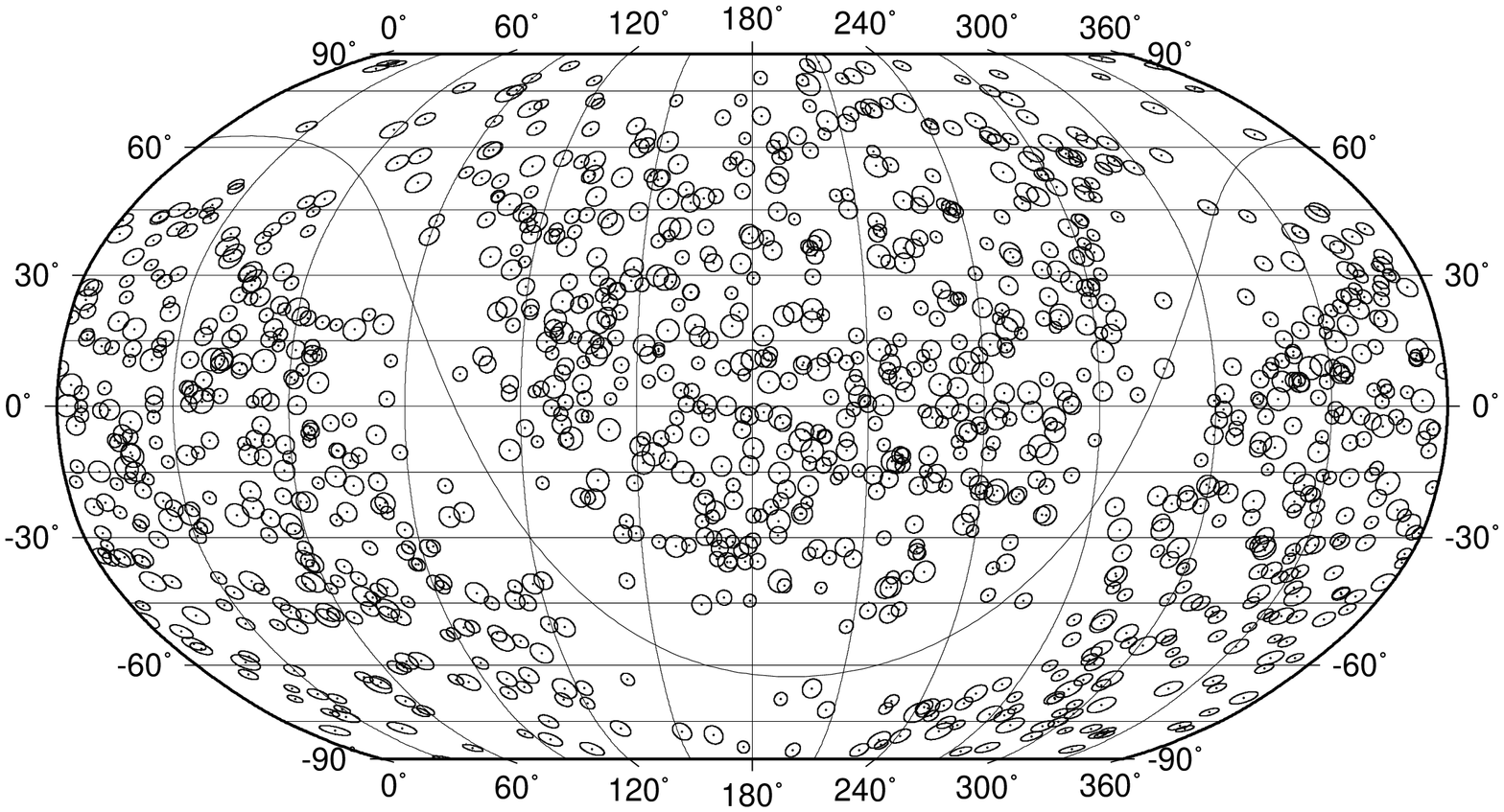}}
}
\hspace{0.03\columnwidth}
\parbox[t]{1.006\columnwidth}{
\tabcolsep=1pt
\begin{tabular}{r|rrrrrrrrrrrrr}
& $a_{1, 1}^{E}$ & $a_{1,-1}^{E}$ & $a_{1, 0}^{E}$ & $a_{2, 1}^{E}$ & $a_{2,-1}^{E}$ &
  $a_{2, 2}^{E}$ & $a_{2,-2}^{E}$ & $a_{2, 0}^{E}$ & $a_{2, 1}^{M}$ & $a_{2,-1}^{M}$ &
  $a_{2, 2}^{M}$ & $a_{2,-2}^{M}$ & $a_{2, 0}^{M}$ \\
\hline
$a_{1, 1}^{E}$ & 1.00 &      &      &      &      &      &      &      &      &      &      &      &      \\
$a_{1,-1}^{E}$ & 0.03 & 1.00 &      &      &      &      &      &      &      &      &      &      &      \\
$a_{1, 0}^{E}$ &-0.09 &-0.01 & 1.00 &      &      &      &      &      &      &      &      &      &      \\
$a_{2, 1}^{E}$ &-0.01 & 0.02 & 0.03 & 1.00 &      &      &      &      &      &      &      &      &      \\
$a_{2,-1}^{E}$ & 0.03 &-0.02 &-0.02 &-0.03 & 1.00 &      &      &      &      &      &      &      &      \\
$a_{2, 2}^{E}$ & 0.03 &-0.03 &-0.02 & 0.01 & 0.06 & 1.00 &      &      &      &      &      &      &      \\
$a_{2,-2}^{E}$ &-0.02 &-0.03 & 0.01 & 0.04 & 0.10 &-0.03 & 1.00 &      &      &      &      &      &      \\
$a_{2, 0}^{E}$ & 0.02 &-0.02 &-0.01 &-0.02 &-0.03 & 0.00 & 0.03 & 1.00 &      &      &      &      &      \\
$a_{2, 1}^{M}$ &-0.06 &-0.03 &-0.08 & 0.00 & 0.04 & 0.00 &-0.01 & 0.02 & 1.00 &      &      &      &      \\
$a_{2,-1}^{M}$ & 0.03 &-0.11 & 0.00 &-0.04 & 0.00 & 0.00 & 0.01 & 0.02 &-0.03 & 1.00 &      &      &      \\
$a_{2, 2}^{M}$ & 0.13 & 0.08 &-0.15 &-0.02 &-0.02 & 0.00 &-0.03 & 0.00 &-0.01 &-0.01 & 1.00 &      &      \\
$a_{2,-2}^{M}$ &-0.09 & 0.04 & 0.07 & 0.01 &-0.03 &-0.03 & 0.01 &-0.02 &-0.02 &-0.04 &-0.05 & 1.00 &      \\
$a_{2, 0}^{M}$ & 0.00 & 0.15 & 0.00 &-0.02 &-0.03 &-0.01 &-0.02 & 0.00 &-0.02 &-0.06 & 0.02 & 0.04 & 1.00 \\
\end{tabular}}
\vskip 2em
\parbox[b]{0.98\columnwidth}{
\raisebox{8em}{\includegraphics[clip,width=0.52\columnwidth,angle=-90]{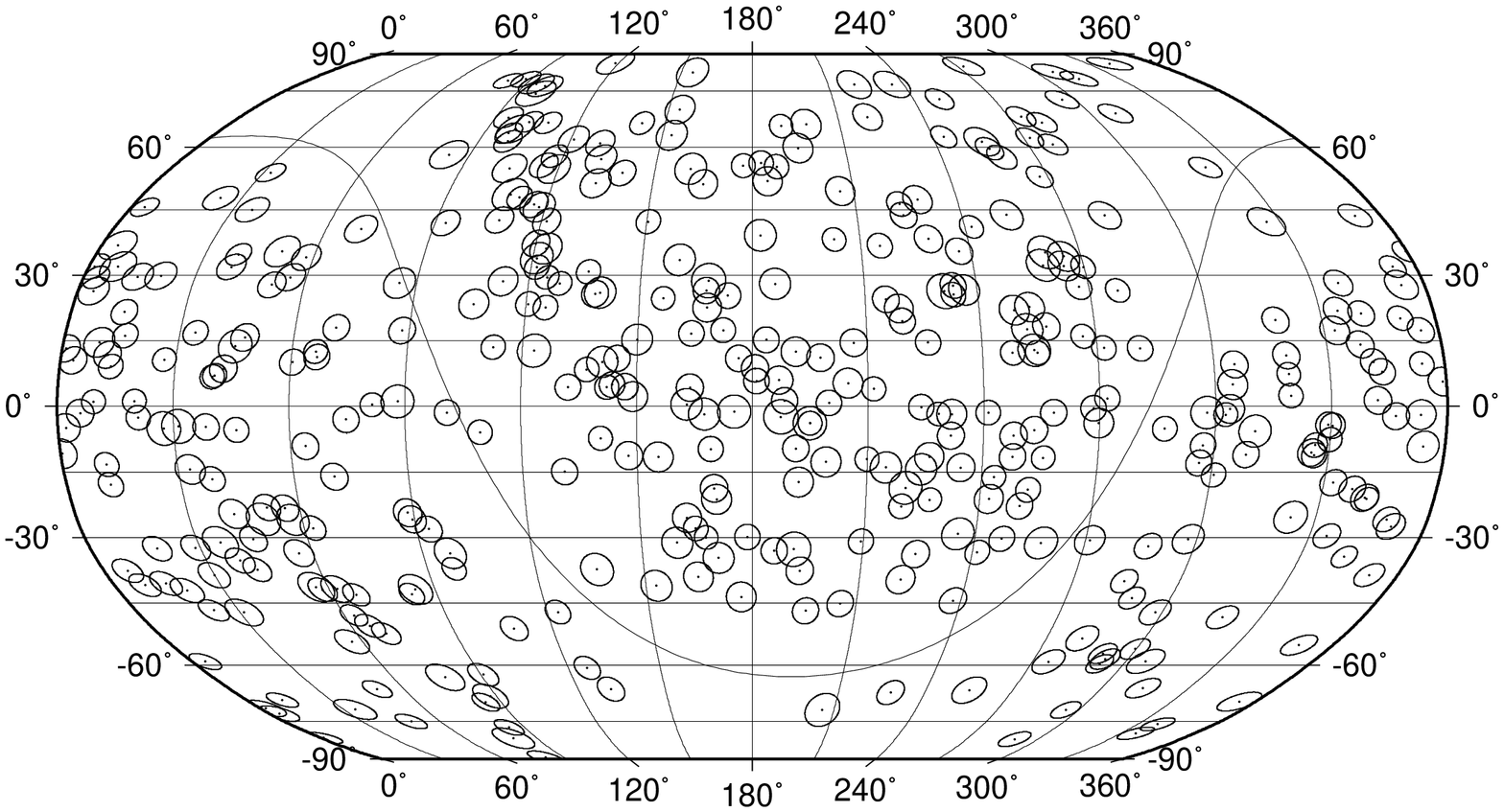}}
}
\hspace{0.03\columnwidth}
\parbox[t]{1.006\columnwidth}{
\tabcolsep=1pt
\begin{tabular}{r|rrrrrrrrrrrrr}
& $a_{1, 1}^{E}$ & $a_{1,-1}^{E}$ & $a_{1, 0}^{E}$ & $a_{2, 1}^{E}$ & $a_{2,-1}^{E}$ &
  $a_{2, 2}^{E}$ & $a_{2,-2}^{E}$ & $a_{2, 0}^{E}$ & $a_{2, 1}^{M}$ & $a_{2,-1}^{M}$ &
  $a_{2, 2}^{M}$ & $a_{2,-2}^{M}$ & $a_{2, 0}^{M}$ \\
\hline
$a_{1, 1}^{E}$ & 1.00 &      &      &      &      &      &      &      &      &      &      &      &      \\
$a_{1,-1}^{E}$ & 0.04 & 1.00 &      &      &      &      &      &      &      &      &      &      &      \\
$a_{1, 0}^{E}$ &-0.09 & 0.01 & 1.00 &      &      &      &      &      &      &      &      &      &      \\
$a_{2, 1}^{E}$ &-0.05 &-0.06 & 0.08 & 1.00 &      &      &      &      &      &      &      &      &      \\
$a_{2,-1}^{E}$ &-0.06 &-0.02 &-0.01 &-0.01 & 1.00 &      &      &      &      &      &      &      &      \\
$a_{2, 2}^{E}$ & 0.01 & 0.01 &-0.06 &-0.01 & 0.02 & 1.00 &      &      &      &      &      &      &      \\
$a_{2,-2}^{E}$ & 0.02 & 0.02 &-0.01 & 0.03 & 0.13 &-0.04 & 1.00 &      &      &      &      &      &      \\
$a_{2, 0}^{E}$ &-0.01 & 0.05 &-0.08 &-0.02 &-0.05 & 0.06 &-0.02 & 1.00 &      &      &      &      &      \\
$a_{2, 1}^{M}$ &-0.05 &-0.04 &-0.07 & 0.00 &-0.03 &-0.09 &-0.03 &-0.02 & 1.00 &      &      &      &      \\
$a_{2,-1}^{M}$ & 0.04 &-0.09 &-0.01 & 0.03 & 0.00 &-0.07 &-0.02 &-0.06 & 0.00 & 1.00 &      &      &      \\
$a_{2, 2}^{M}$ & 0.18 & 0.04 &-0.13 &-0.04 & 0.00 & 0.03 & 0.03 & 0.02 & 0.02 & 0.02 & 1.00 &      &      \\
$a_{2,-2}^{M}$ &-0.03 & 0.09 & 0.08 & 0.00 & 0.00 & 0.02 &-0.02 &-0.06 &-0.01 &-0.06 &-0.05 & 1.00 &      \\
$a_{2, 0}^{M}$ & 0.03 & 0.15 &-0.01 & 0.01 & 0.04 & 0.02 &-0.05 & 0.01 &-0.02 &-0.06 & 0.06 & 0.00 & 1.00 \\
\end{tabular}}
\vskip 2em
\parbox[b]{0.98\columnwidth}{
\raisebox{8em}{\includegraphics[clip,width=0.52\columnwidth,angle=-90]{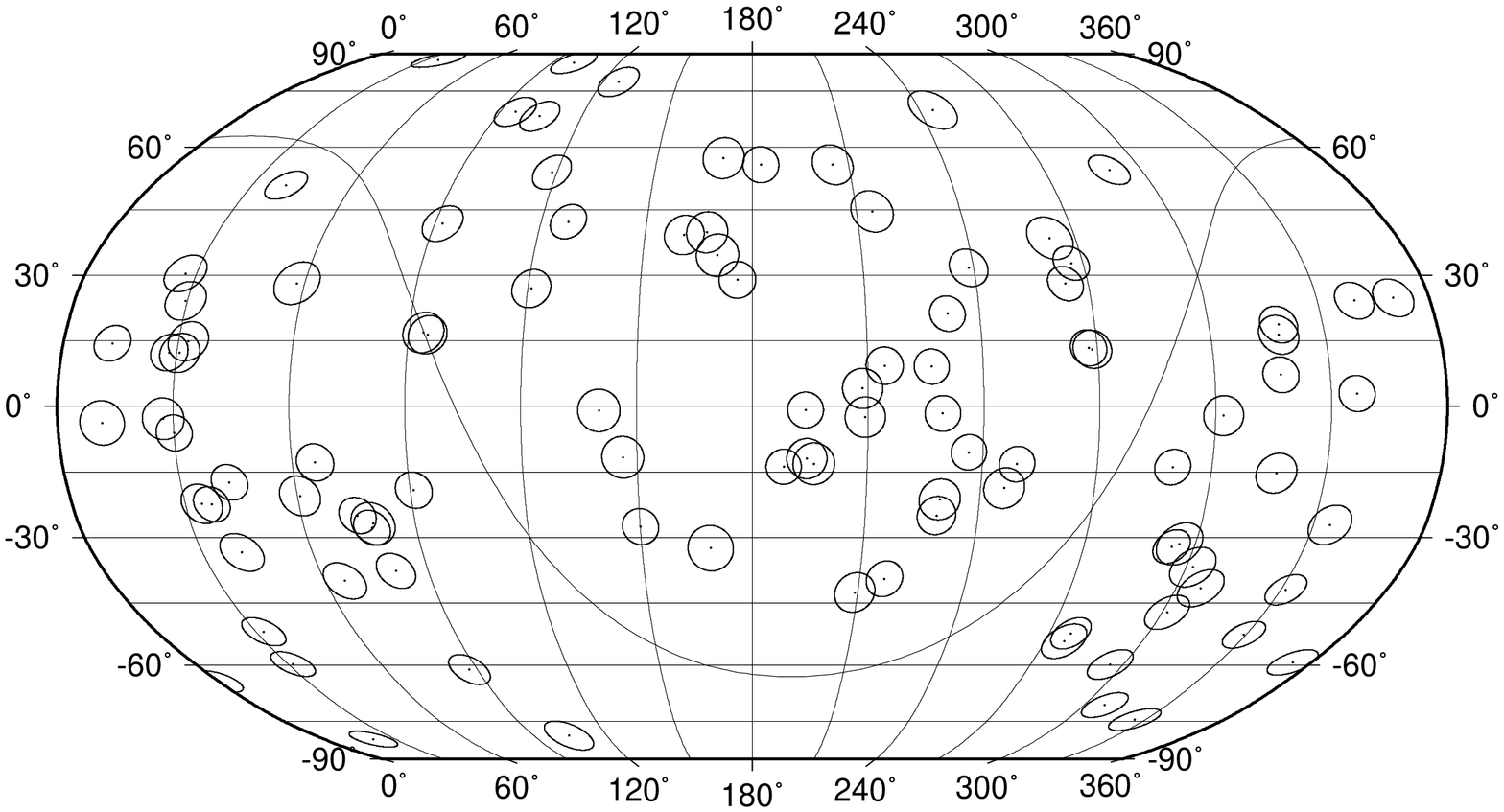}}
}
\hspace{0.03\columnwidth}
\parbox[t]{1.006\columnwidth}{
\tabcolsep=1pt
\begin{tabular}{r|rrrrrrrrrrrrr}
& $a_{1, 1}^{E}$ & $a_{1,-1}^{E}$ & $a_{1, 0}^{E}$ & $a_{2, 1}^{E}$ & $a_{2,-1}^{E}$ &
  $a_{2, 2}^{E}$ & $a_{2,-2}^{E}$ & $a_{2, 0}^{E}$ & $a_{2, 1}^{M}$ & $a_{2,-1}^{M}$ &
  $a_{2, 2}^{M}$ & $a_{2,-2}^{M}$ & $a_{2, 0}^{M}$ \\
\hline
$a_{1, 1}^{E}$ & 1.00 &      &      &      &      &      &      &      &      &      &      &      &      \\
$a_{1,-1}^{E}$ & 0.11 & 1.00 &      &      &      &      &      &      &      &      &      &      &      \\
$a_{1, 0}^{E}$ &-0.11 & 0.03 & 1.00 &      &      &      &      &      &      &      &      &      &      \\
$a_{2, 1}^{E}$ &-0.04 &-0.03 & 0.09 & 1.00 &      &      &      &      &      &      &      &      &      \\
$a_{2,-1}^{E}$ & 0.03 &-0.03 &-0.05 &-0.11 & 1.00 &      &      &      &      &      &      &      &      \\
$a_{2, 2}^{E}$ & 0.04 &-0.04 & 0.00 & 0.03 & 0.00 & 1.00 &      &      &      &      &      &      &      \\
$a_{2,-2}^{E}$ &-0.06 &-0.06 & 0.04 & 0.08 & 0.19 & 0.01 & 1.00 &      &      &      &      &      &      \\
$a_{2, 0}^{E}$ & 0.02 &-0.02 & 0.06 & 0.09 &-0.14 & 0.02 &-0.08 & 1.00 &      &      &      &      &      \\
$a_{2, 1}^{M}$ & 0.00 &-0.09 &-0.11 & 0.00 &-0.01 &-0.02 & 0.06 & 0.04 & 1.00 &      &      &      &      \\
$a_{2,-1}^{M}$ & 0.10 &-0.03 &-0.03 & 0.03 & 0.00 & 0.09 &-0.05 & 0.02 &-0.12 & 1.00 &      &      &      \\
$a_{2, 2}^{M}$ & 0.19 & 0.11 &-0.03 &-0.05 &-0.03 & 0.01 & 0.07 &-0.06 & 0.01 &-0.01 & 1.00 &      &      \\
$a_{2,-2}^{M}$ &-0.07 & 0.07 & 0.18 & 0.03 &-0.06 & 0.09 &-0.01 & 0.00 &-0.03 &-0.01 & 0.01 & 1.00 &      \\
$a_{2, 0}^{M}$ & 0.08 & 0.21 &-0.02 &-0.06 &-0.02 &-0.09 &-0.03 &-0.01 & 0.07 &-0.13 & 0.06 &-0.05 & 1.00 \\
\end{tabular}}
}
\caption{Sky coverage and correlation matrix for 13 vector spherical harmonics of first and second
order for modelled source distribution and four redshift intervals.
From top to bottom: $0<z<1$, $1<z<2$, $2<z<3$, $3<z<4$.
The circle size corresponds to the redshift value.}
\label{fig:distr-corr_z_model}
\end{figure*}

\begin{figure*}
\centering
{\tiny
\parbox[b]{0.98\columnwidth}{
\raisebox{8em}{\includegraphics[clip,width=0.52\columnwidth,angle=-90]{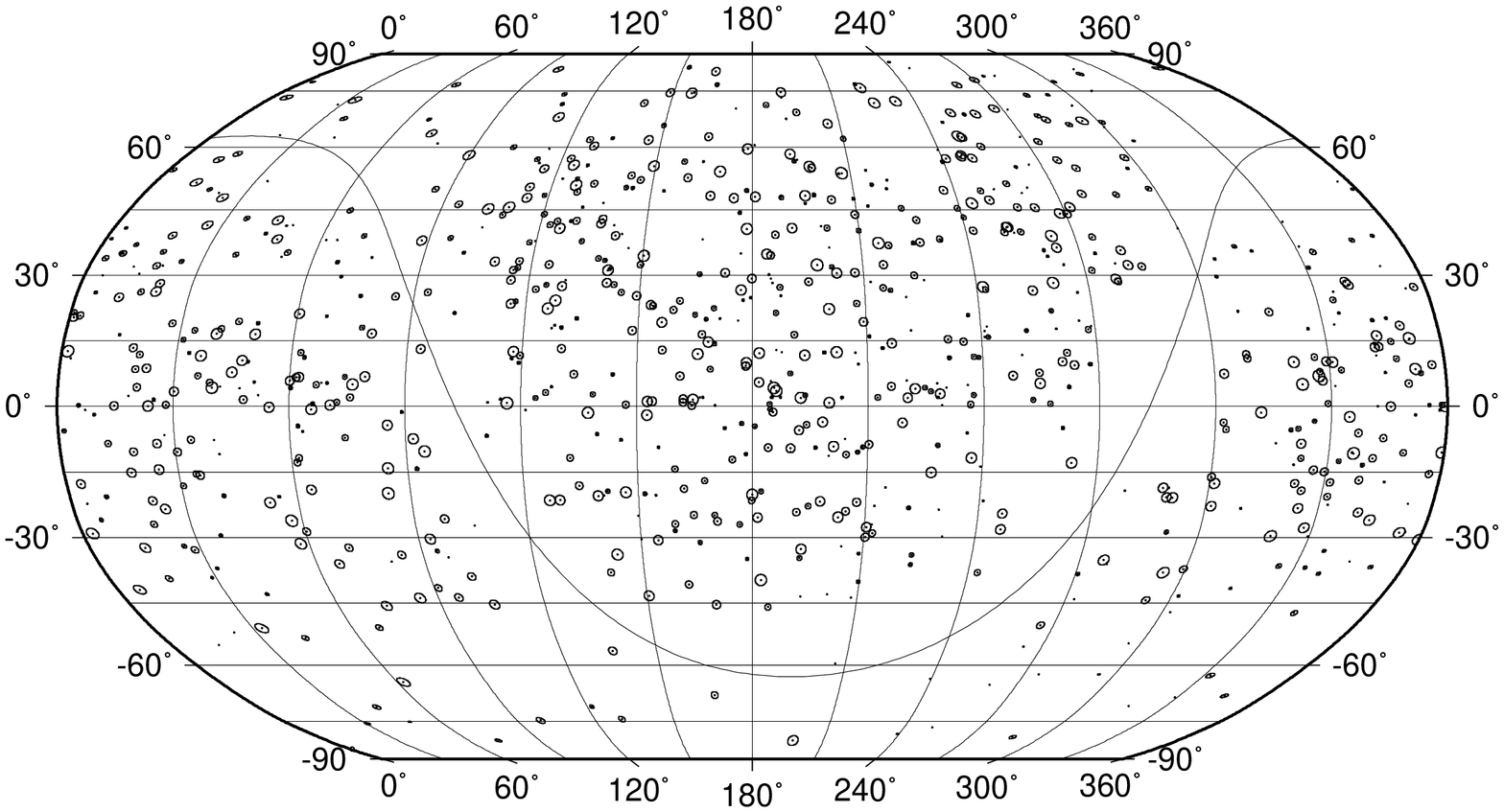}}
}
\hspace{0.03\columnwidth}
\parbox[t]{1.006\columnwidth}{
\tabcolsep=1pt
\begin{tabular}{r|rrrrrrrrrrrrr}
& $a_{1, 1}^{E}$ & $a_{1,-1}^{E}$ & $a_{1, 0}^{E}$ & $a_{2, 1}^{E}$ & $a_{2,-1}^{E}$ &
  $a_{2, 2}^{E}$ & $a_{2,-2}^{E}$ & $a_{2, 0}^{E}$ & $a_{2, 1}^{M}$ & $a_{2,-1}^{M}$ &
  $a_{2, 2}^{M}$ & $a_{2,-2}^{M}$ & $a_{2, 0}^{M}$ \\
\hline
$a_{1, 1}^{E}$ & 1.00 &      &      &      &      &      &      &      &      &      &      &      &      \\
$a_{1,-1}^{E}$ & 0.01 & 1.00 &      &      &      &      &      &      &      &      &      &      &      \\
$a_{1, 0}^{E}$ &-0.08 & 0.00 & 1.00 &      &      &      &      &      &      &      &      &      &      \\
$a_{2, 1}^{E}$ & 0.03 &-0.24 &-0.03 & 1.00 &      &      &      &      &      &      &      &      &      \\
$a_{2,-1}^{E}$ &-0.22 & 0.01 & 0.02 &-0.04 & 1.00 &      &      &      &      &      &      &      &      \\
$a_{2, 2}^{E}$ &-0.02 & 0.11 & 0.01 &-0.04 & 0.05 & 1.00 &      &      &      &      &      &      &      \\
$a_{2,-2}^{E}$ & 0.07 &-0.05 &-0.04 & 0.04 & 0.07 &-0.05 & 1.00 &      &      &      &      &      &      \\
$a_{2, 0}^{E}$ &-0.07 &-0.02 &-0.27 &-0.01 & 0.02 & 0.00 & 0.06 & 1.00 &      &      &      &      &      \\
$a_{2, 1}^{M}$ & 0.00 &-0.03 &-0.05 & 0.01 &-0.09 & 0.02 & 0.01 &-0.10 & 1.00 &      &      &      &      \\
$a_{2,-1}^{M}$ & 0.03 &-0.23 &-0.01 & 0.15 &-0.01 &-0.09 & 0.05 & 0.04 &-0.03 & 1.00 &      &      &      \\
$a_{2, 2}^{M}$ & 0.13 & 0.06 &-0.21 & 0.05 & 0.05 & 0.01 & 0.02 & 0.06 &-0.02 &-0.02 & 1.00 &      &      \\
$a_{2,-2}^{M}$ &-0.08 & 0.08 & 0.07 &-0.06 &-0.04 & 0.01 &-0.03 & 0.00 &-0.03 &-0.07 &-0.06 & 1.00 &      \\
$a_{2, 0}^{M}$ &-0.01 & 0.13 & 0.00 & 0.09 &-0.03 & 0.03 & 0.00 & 0.00 &-0.03 &-0.03 & 0.02 & 0.07 & 1.00 \\
\end{tabular}}
\vskip 2em
\parbox[b]{0.98\columnwidth}{
\raisebox{8em}{\includegraphics[clip,width=0.52\columnwidth,angle=-90]{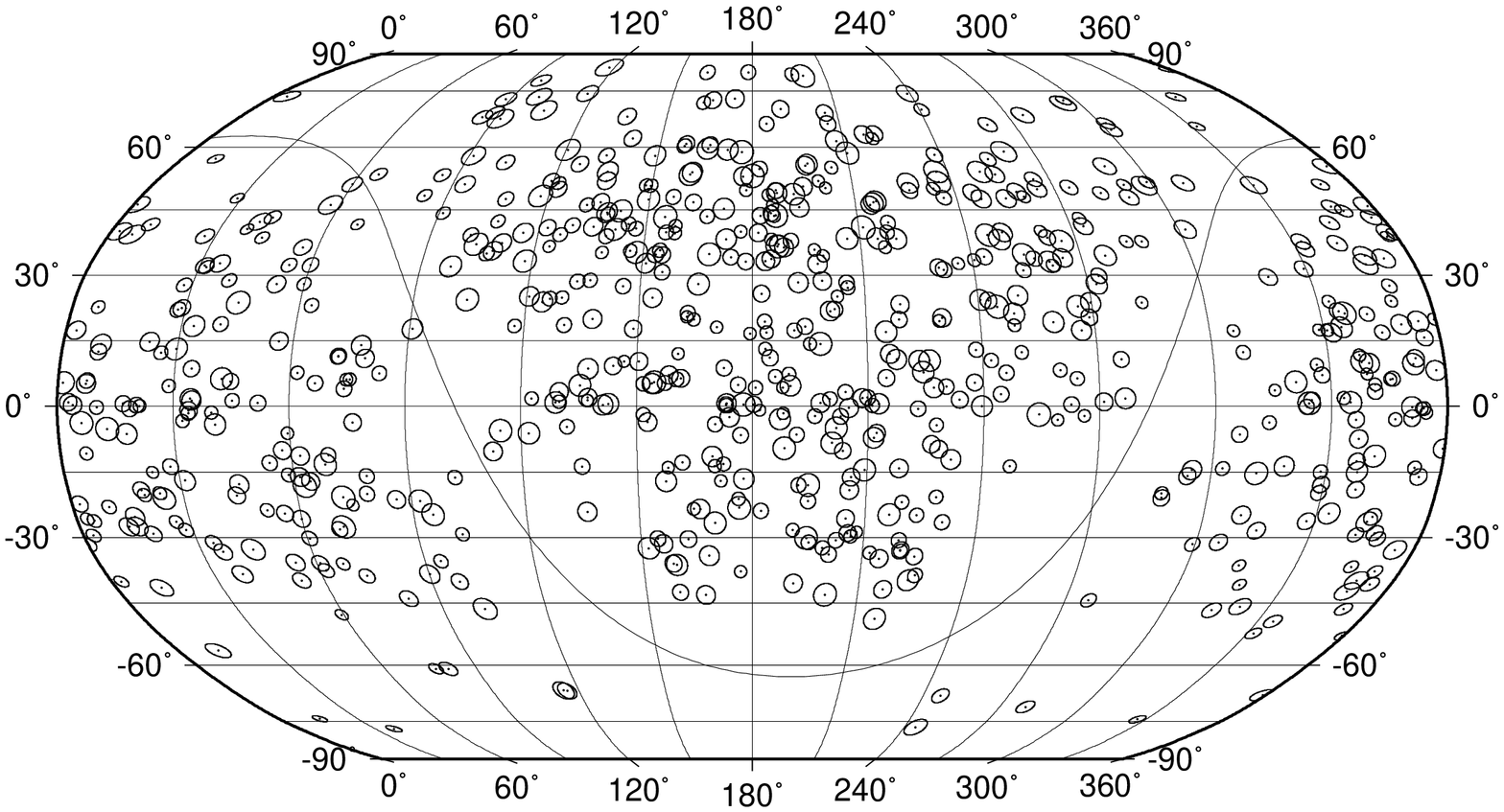}}
}
\hspace{0.03\columnwidth}
\parbox[t]{1.006\columnwidth}{
\tabcolsep=1pt
\begin{tabular}{r|rrrrrrrrrrrrr}
& $a_{1, 1}^{E}$ & $a_{1,-1}^{E}$ & $a_{1, 0}^{E}$ & $a_{2, 1}^{E}$ & $a_{2,-1}^{E}$ &
  $a_{2, 2}^{E}$ & $a_{2,-2}^{E}$ & $a_{2, 0}^{E}$ & $a_{2, 1}^{M}$ & $a_{2,-1}^{M}$ &
  $a_{2, 2}^{M}$ & $a_{2,-2}^{M}$ & $a_{2, 0}^{M}$ \\
\hline
$a_{1, 1}^{E}$ & 1.00 &      &      &      &      &      &     &       &      &      &      &      &      \\
$a_{1,-1}^{E}$ & 0.03 & 1.00 &      &      &      &      &     &       &      &      &      &      &      \\
$a_{1, 0}^{E}$ &-0.12 &-0.05 & 1.00 &      &      &      &     &       &      &      &      &      &      \\
$a_{2, 1}^{E}$ &-0.02 &-0.25 & 0.01 & 1.00 &      &      &     &       &      &      &      &      &      \\
$a_{2,-1}^{E}$ &-0.22 &-0.01 & 0.02 &-0.03 & 1.00 &      &     &       &      &      &      &      &      \\
$a_{2, 2}^{E}$ &-0.02 & 0.13 &-0.05 &-0.02 & 0.06 & 1.00 &     &       &      &      &      &      &      \\
$a_{2,-2}^{E}$ & 0.09 &-0.03 &-0.06 & 0.05 & 0.09 &-0.01 & 1.00&       &      &      &      &      &      \\
$a_{2, 0}^{E}$ &-0.07 & 0.00 &-0.24 & 0.04 & 0.04 & 0.04 & 0.15&  1.00 &      &      &      &      &      \\
$a_{2, 1}^{M}$ & 0.00 &-0.03 &-0.08 & 0.01 &-0.10 & 0.02 & 0.02& -0.13 & 1.00 &      &      &      &      \\
$a_{2,-1}^{M}$ & 0.04 &-0.33 & 0.05 & 0.17 &-0.01 &-0.11 & 0.00&  0.00 &-0.04 & 1.00 &      &      &      \\
$a_{2, 2}^{M}$ & 0.12 & 0.09 &-0.28 &-0.02 & 0.06 & 0.00 & 0.05&  0.07 &-0.06 &-0.06 & 1.00 &      &      \\
$a_{2,-2}^{M}$ &-0.11 & 0.06 & 0.10 &-0.06 & 0.01 & 0.03 &-0.01& -0.06 & 0.03 &-0.08 &-0.05 & 1.00 &      \\
$a_{2, 0}^{M}$ &-0.07 & 0.19 & 0.00 & 0.09 & 0.00 & 0.03 &-0.05& -0.01 & 0.02 &-0.07 & 0.03 & 0.16 & 1.00 \\
\end{tabular}}
\vskip 2em
\parbox[b]{0.98\columnwidth}{
\raisebox{8em}{\includegraphics[clip,width=0.52\columnwidth,angle=-90]{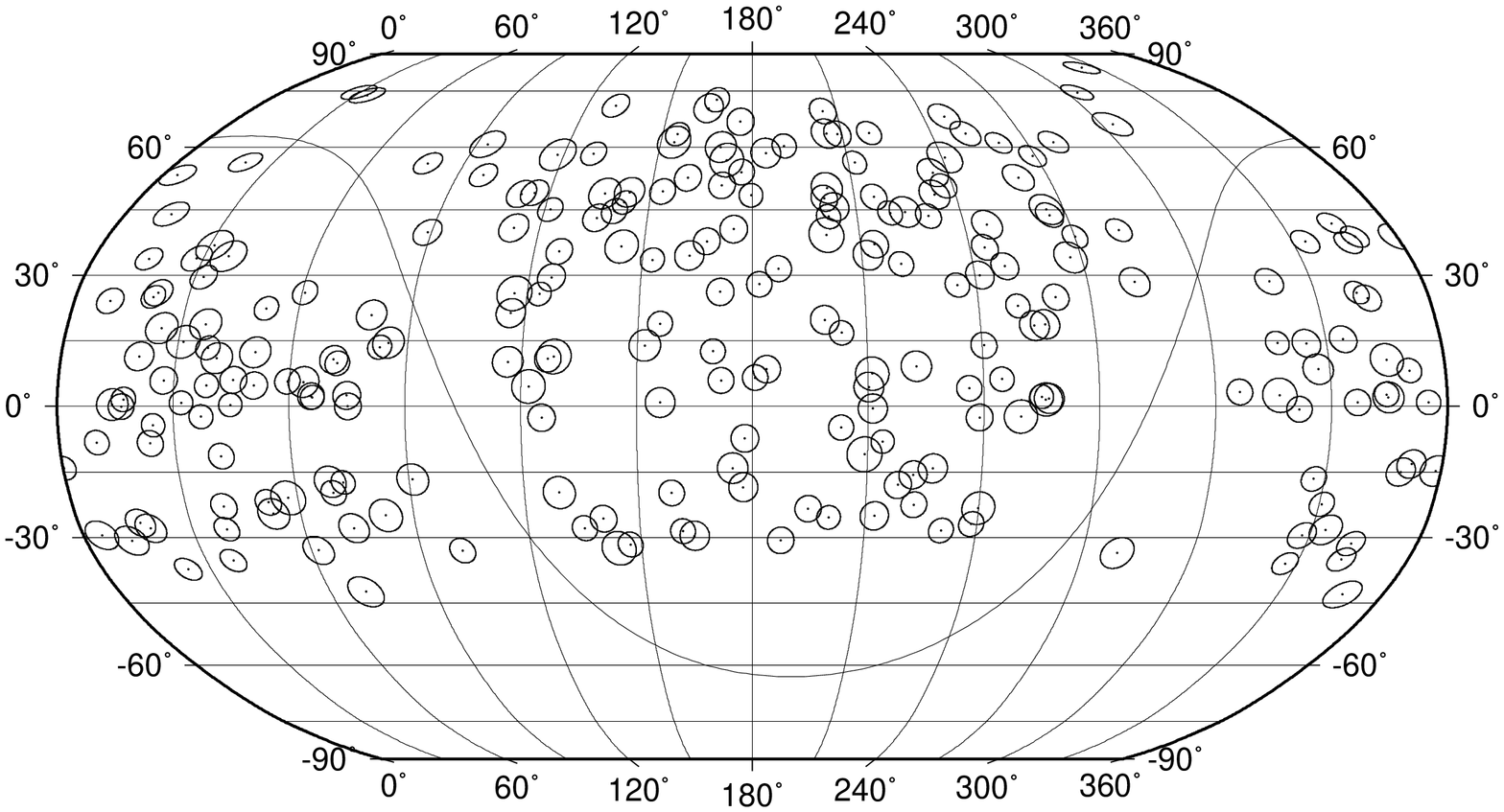}}
}
\hspace{0.03\columnwidth}
\parbox[t]{1.006\columnwidth}{
\tabcolsep=1pt
\begin{tabular}{r|rrrrrrrrrrrrr}
& $a_{1, 1}^{E}$ & $a_{1,-1}^{E}$ & $a_{1, 0}^{E}$ & $a_{2, 1}^{E}$ & $a_{2,-1}^{E}$ &
  $a_{2, 2}^{E}$ & $a_{2,-2}^{E}$ & $a_{2, 0}^{E}$ & $a_{2, 1}^{M}$ & $a_{2,-1}^{M}$ &
  $a_{2, 2}^{M}$ & $a_{2,-2}^{M}$ & $a_{2, 0}^{M}$ \\
\hline
$a_{1, 1}^{E}$ & 1.00 &      &      &      &      &      &      &      &      &      &      &      &      \\
$a_{1,-1}^{E}$ & 0.01 & 1.00 &      &      &      &      &      &      &      &      &      &      &      \\
$a_{1, 0}^{E}$ &-0.08 &-0.01 & 1.00 &      &      &      &      &      &      &      &      &      &      \\
$a_{2, 1}^{E}$ & 0.06 &-0.34 &-0.09 & 1.00 &      &      &      &      &      &      &      &      &      \\
$a_{2,-1}^{E}$ &-0.34 & 0.06 &-0.07 &-0.13 & 1.00 &      &      &      &      &      &      &      &      \\
$a_{2, 2}^{E}$ &-0.09 & 0.12 &-0.03 &-0.02 & 0.09 & 1.00 &      &      &      &      &      &      &      \\
$a_{2,-2}^{E}$ & 0.06 &-0.01 &-0.07 & 0.03 & 0.09 &-0.03 & 1.00 &      &      &      &      &      &      \\
$a_{2, 0}^{E}$ &-0.12 & 0.01 &-0.29 & 0.03 & 0.05 & 0.08 & 0.12 & 1.00 &      &      &      &      &      \\
$a_{2, 1}^{M}$ & 0.13 &-0.10 &-0.07 & 0.05 &-0.19 & 0.01 & 0.05 &-0.15 & 1.00 &      &      &      &      \\
$a_{2,-1}^{M}$ & 0.14 &-0.33 & 0.01 & 0.25 &-0.08 &-0.25 & 0.08 & 0.00 &-0.05 & 1.00 &      &      &      \\
$a_{2, 2}^{M}$ & 0.13 & 0.06 &-0.19 & 0.08 & 0.00 & 0.05 & 0.03 & 0.05 & 0.00 &-0.03 & 1.00 &      &      \\
$a_{2,-2}^{M}$ &-0.13 & 0.05 & 0.20 &-0.03 & 0.04 & 0.03 &-0.05 &-0.04 &-0.03 &-0.09 &-0.07 & 1.00 &      \\
$a_{2, 0}^{M}$ &-0.07 & 0.20 & 0.00 & 0.09 & 0.01 & 0.05 & 0.00 & 0.01 &-0.02 &-0.12 & 0.08 & 0.14 & 1.00 \\
\end{tabular}}
\vskip 2em
\parbox[b]{0.98\columnwidth}{
\raisebox{8em}{\includegraphics[clip,width=0.52\columnwidth,angle=-90]{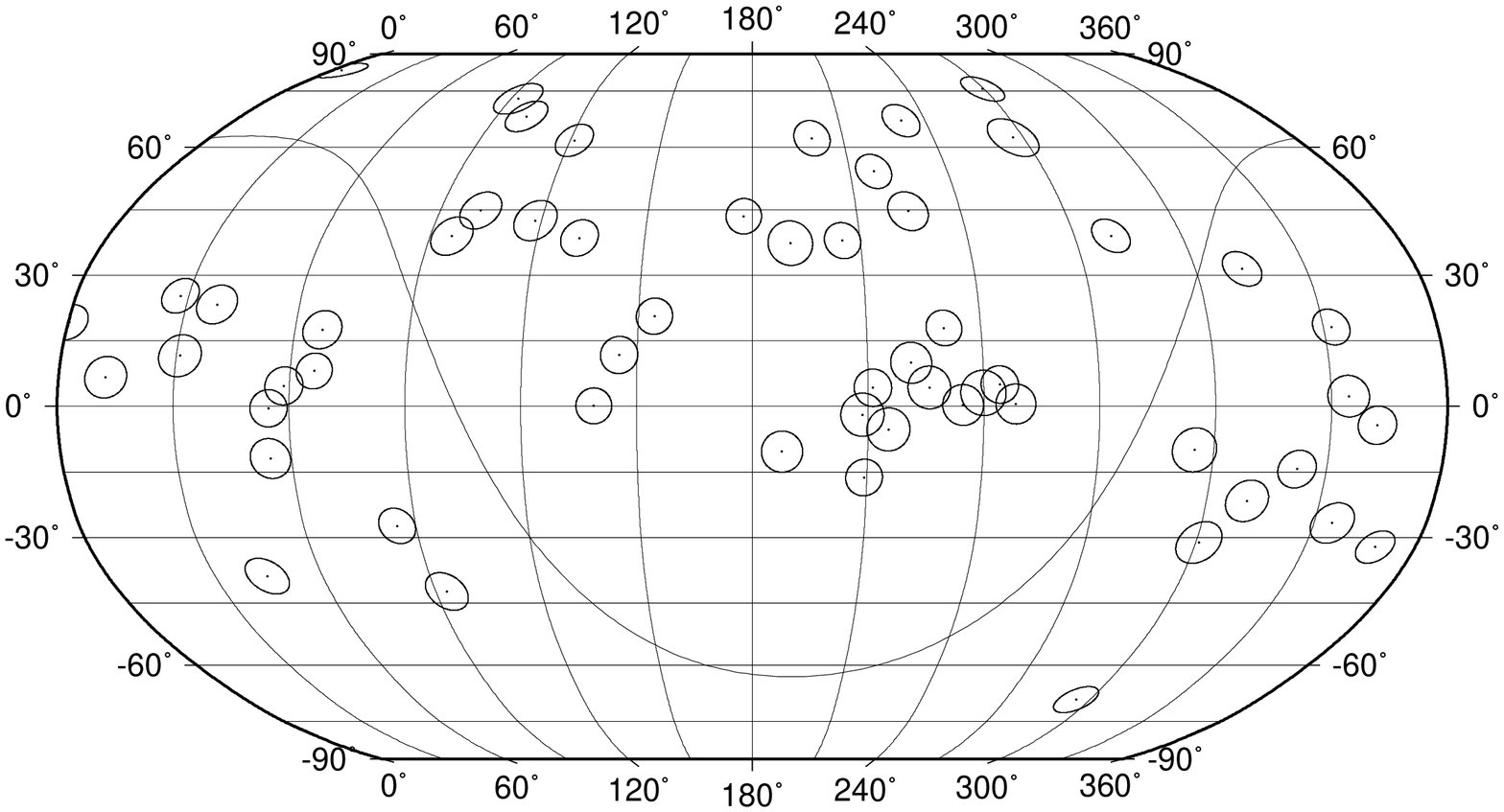}}
}
\hspace{0.03\columnwidth}
\parbox[t]{1.006\columnwidth}{
\tabcolsep=1pt
\begin{tabular}{r|rrrrrrrrrrrrr}
& $a_{1, 1}^{E}$ & $a_{1,-1}^{E}$ & $a_{1, 0}^{E}$ & $a_{2, 1}^{E}$ & $a_{2,-1}^{E}$ &
  $a_{2, 2}^{E}$ & $a_{2,-2}^{E}$ & $a_{2, 0}^{E}$ & $a_{2, 1}^{M}$ & $a_{2,-1}^{M}$ &
  $a_{2, 2}^{M}$ & $a_{2,-2}^{M}$ & $a_{2, 0}^{M}$ \\
\hline
$a_{1, 1}^{E}$ & 1.00 &      &      &      &      &      &      &      &      &      &      &      &      \\
$a_{1,-1}^{E}$ &-0.02 & 1.00 &      &      &      &      &      &      &      &      &      &      &      \\
$a_{1, 0}^{E}$ &-0.13 & 0.08 & 1.00 &      &      &      &      &      &      &      &      &      &      \\
$a_{2, 1}^{E}$ & 0.22 &-0.51 &-0.14 & 1.00 &      &      &      &      &      &      &      &      &      \\
$a_{2,-1}^{E}$ &-0.38 & 0.31 & 0.31 &-0.43 & 1.00 &      &      &      &      &      &      &      &      \\
$a_{2, 2}^{E}$ &-0.19 & 0.33 & 0.03 &-0.22 & 0.24 & 1.00 &      &      &      &      &      &      &      \\
$a_{2,-2}^{E}$ & 0.19 & 0.08 & 0.07 & 0.01 & 0.09 &-0.10 & 1.00 &      &      &      &      &      &      \\
$a_{2, 0}^{E}$ &-0.24 & 0.22 &-0.39 &-0.18 & 0.04 & 0.25 &-0.17 & 1.00 &      &      &      &      &      \\
$a_{2, 1}^{M}$ & 0.20 &-0.40 &-0.12 & 0.24 &-0.34 &-0.18 &-0.02 &-0.25 & 1.00 &      &      &      &      \\
$a_{2,-1}^{M}$ & 0.34 &-0.20 &-0.05 & 0.31 &-0.19 &-0.19 &-0.03 &-0.19 &-0.06 & 1.00 &      &      &      \\
$a_{2, 2}^{M}$ & 0.19 & 0.10 & 0.05 &-0.02 & 0.18 & 0.12 & 0.05 &-0.09 &-0.05 & 0.13 & 1.00 &      &      \\
$a_{2,-2}^{M}$ &-0.18 & 0.19 & 0.49 &-0.31 & 0.28 & 0.07 &-0.05 &-0.06 &-0.09 &-0.19 &-0.08 & 1.00 &      \\
$a_{2, 0}^{M}$ & 0.03 & 0.24 &-0.01 & 0.06 & 0.14 & 0.06 & 0.11 & 0.03 &-0.15 &-0.02 & 0.20 &-0.11 & 1.00 \\
\end{tabular}}
}
\caption{Sky coverage and correlation matrix for 13 vector spherical harmonics of first and second
order for real source distribution and four redshift intervals.
From top to bottom: $0<z<1$, $1<z<2$, $2<z<3$, $3<z<4$.
The circle size corresponds to the redshift value.}
\label{fig:distr-corr_z_real}
\end{figure*}

A comparison of the correlation matrices corresponding to two modelled
distribution shown in Fig~\ref{fig:distr-corr_all} highlights that the avoidance
zone increases the maximum correlation (in absolute values) between the
estimated spherical harmonics from 0.03 to 0.16. It is
obvious that the paucity of the radio sources under declination $-30^\circ$
(for the real distribution) results in further
increase of the correlation to the maximum absolute value 0.28.

Once the separation on different zones on redshift is applicable, we
also considered the correlation between observables in four zones for
the both model and real distribution (Figs.~\ref{fig:distr-corr_z_model}
and~\ref{fig:distr-corr_z_real}, respectively).
The model distribution without essential decrease in the number of
sources around the South Celestial Pole produces almost equal maximum
correlation parameter for all four redshift zones (from 0.15 to 0.19)

In opposite, a deficit of the radio sources around the South Celestial Pole
is more crucial. The data presented in Figs.~\ref{fig:distr-corr_z_real}
demonstrate the impact of the North-South asymmetry in the real
distribution of the radio sources. For the zone of redshift between
3 and 4 the maximum correlation reaches 0.51.

As we learnt from the comparison of correlation matrices for the
limited number of radio sources unevenly distributed around the
celestial sphere, the correlation between estimated spherical
harmonics might increase dramatically to level of 0.8--0.9,
especially of not all the sources are observed.  Whereas, the
individual apparent motion due to intrinsic structure reaches
500~$\mu$as/yr (Feissel-Vernier 2003; MacMillan \& Ma 2007;
Titov 2007), they could propagate systematically to the estimated parameters,
if the number of objects is not sufficient, i.e. for the case of high redshift.

\section{Conclusion}

To conclude, it is necessary to note that this `historic' deficit of
the radio sources (and, additionally, the radio sources with measured
redshift) might cause problems with further investigation of the
hardly detectable systematic effects in the proper motion of the
reference radio sources. Therefore, large observational projects for
spectroscopy of the astrometric radio sources in the Southern
hemisphere are very important.  Nonetheless, some observations in the
North hemisphere also need to be undertaken.

Independently, the MASIV scintillation survey
(Lovell et al. 2003, 2009) also demonstrates a highly significant dramatic
decrease in the numbers of scintillators for redshifts in excess of z = 2.
The lack of scintillation at high redshifts is clear evidence for an
increase in the source angular sizes with increasing redshift.
Such an increase may be cosmological in origin or may be a propagation
effect of inter-galactic scattering (Lovell et al. 2009).

To observe and study radio source in both frequency range (optical and
radio) characteristics such as visual magnitude, redshift in optic and
flux density in several radio bands have to be measured.
We also need to be sure that the same source is observed
by the optical and radio instruments. Due to possible misalignment
between optical and radio positions the physical characteristics might
help to solve the problem of identification.

To chase this aim, a new list of optical characteristics of 4261 astrometric
radio sources, OCARS, including all 717 ICRF-Ext.2 sources has been compiled.
The OCARS list includes source type, redshift and visual magnitude (when available).
Detailed comments are provided when necessary, which is especially useful
in understanding of incomplete, contradictory and controversial
astrophysical data.
The OCARS may serve to various VLBI tasks, for instance:

\begin{itemize}
\item As a supplement material for the second ICRF realization ICRF2 (Ma 2008).
\item As a database for VLBI data analysis.
\item For planning of IVS and other observing programs, in order to enrich the
observational history of the sources with reliable determined redshift.
\item For future link between optical (GAIA) and radio (ICRF) celestial
reference frames, once the GAIA optical position of about 100,000 quasars will be available.
\end{itemize}

We performed a detailed comparison of the OCARS list with the official IERS list,
and found many discrepancies for about a half of common sources.  This
comparison showed that the IERS list seems to be outdated and should be
used with care.
Besides, the IERS list, being intended to provide physical characteristics
of the IERS sources only, contains only a small fraction of the whole set
of astrometric radio sources used nowadays.

We also compared the OCARS with the newest Large Quasar Astrometric Catalogue LQAC,
and found discrepancies which worth further investigating.
Most of discrepancies seems to be a result of different object identification.

This is only the first stage of our work. We are planning the following steps:
\begin{itemize}
\item To continue searching for the missing and checking out the
ambiguous characteristics through literature and astronomical
databases.
\item To organize photometric and spectroscopy observations of astrometric
radio sources with large optical telescopes.  In particular, such
an observational program started at Pulkovo Observatory in 2008.
Observations are being made on the 6-m BTA telescope of the Special
Astrophysical Observatory in North Caucasus.
\end{itemize}

The list of optical characteristics of astrometric
radio sources presented in this paper is publicly available at
\mbox{\tt www.gao.spb.ru/english/as/ac\_vlbi/sou\_car.dat}
and is updated once in several months.

\section*{Acknowledgements}
This research has made use of the NASA/IPAC Extragalactic Data\-base (NED)
which is operated by the Jet Propulsion Laboratory, California Institute
of Technology, under contract with the National Aeronautics and Space
Administration, the SIMBAD and HyperLeda databases,
and the CfA-Arizona Space Telescope LEns Survey (CASTLES).
Valuable comments and suggestions of the anonymous reviewer are highly appreciated.

\end{document}